\documentclass[12pt]{article}
\usepackage{cmbright}
\usepackage[OT1]{fontenc}

\usepackage{amssymb}
\usepackage{amsfonts}
\usepackage{amsmath}
\usepackage{parskip} 
\usepackage[left=1in, right=1in, top=1in, bottom=1in]{geometry}
\usepackage{amssymb}
\usepackage{amsmath}
\usepackage{soul}
\usepackage{epigraph}
\usepackage{caption}
\usepackage{subcaption}
\usepackage{tikzsymbols}
\usepackage{booktabs}
\usepackage[colorinlistoftodos,prependcaption,textsize=tiny,linecolor=black,backgroundcolor=lightgray!25]{todonotes}

\definecolor{osu}{RGB}{215, 63, 9}
\definecolor{fruitpushorange}{RGB}{255, 127, 0}
\definecolor{gadflyblue}{RGB}{0, 190, 255}
\definecolor{glassblue}{RGB}{165, 241, 246}
\definecolor{gadflyyellow}{RGB}{212, 202, 58}
\definecolor{gadflygreen}{RGB}{103, 225, 181}
\definecolor{gadflypink}{RGB}{235, 172, 250}
\definecolor{icolor}{rgb}{1.000000, 0.403746, 0.397903}
\definecolor{cool_blue}{RGB}{24, 132, 193}
\definecolor{cool_green}{RGB}{62, 182, 159}
\definecolor{cool_gray}{RGB}{224, 229, 231}
\definecolor{c1}{rgb}{0.12, 0.56, 1.0}

\usepackage{amsmath,amssymb}

\usepackage{cite}
\usepackage[rightcaption]{sidecap}
\usepackage{hyperref}
\usepackage[font=small, justification=centering, width=0.95\textwidth]{caption}
\tikzstyle{block} = [draw, 
    fill=cool_gray!10,
    rectangle, line width=0.5mm,
    minimum height=4.75em, minimum width=4.75em]

\tikzstyle{mybox} = [draw=black, fill=cool_gray!10, very thick,
    rectangle, rounded corners, inner sep=5pt, inner ysep=5pt]
\tikzstyle{sum} = [draw, fill=ivory!20, circle, node distance=1cm, line width=0.5mm]
\tikzstyle{input} = [coordinate]
\tikzstyle{output} = [coordinate]
\usepackage{tikz}
\usepackage{tcolorbox} 
\tcbuselibrary{breakable}
\usetikzlibrary{patterns}
\newtcolorbox[auto counter
]{mybox}[2][]{
title=#2,#1
}

\newcommand\Epsilon{E}
\newcommand\thedata {$\{(t_i,\theta_{i, \text{obs}})\}_{i=1}^{N}$\xspace}
\newcommand\thedatatr {$(t^\prime, \theta_{\text{obs}}^\prime)$\xspace}

\newcommand \rocketemoji{\includegraphics[height=1.6\fontcharht\font`A]{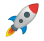} \xspace}
\newcommand \bookemoji{\includegraphics[height=1.6\fontcharht\font`A]{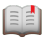} \xspace}
\newcommand \lightbulbemoji{\includegraphics[height=1.6\fontcharht\font`A]{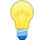} \xspace}

\DeclareFontFamily{OT1}{cmbr}{\hyphenchar\font45 }
\DeclareFontShape{OT1}{cmbr}{m}{n}{%
  <-9>cmbr8
  <9-10>cmbr9
  <10-17>cmbr10
  <17->cmbr17
}{}
\DeclareFontShape{OT1}{cmbr}{m}{sl}{%
  <-9>cmbrsl8
  <9-10>cmbrsl9
  <10-17>cmbrsl10
  <17->cmbrsl17
}{}
\DeclareFontShape{OT1}{cmbr}{m}{it}{%
  <->ssub*cmbr/m/sl
}{}
\DeclareFontShape{OT1}{cmbr}{b}{n}{%
  <->ssub*cmbr/bx/n
}{}
\DeclareFontShape{OT1}{cmbr}{bx}{n}{%
  <->cmbrbx10
}{}

\newcommand{\highlight}[2][red!50]{\mathpalette{\highlightwithstyle[#1]}{#2}}
\newcommand{\highlightwithstyle}[3][red!50]{
  \begingroup                         
    \sbox0{$\mathsurround 0pt #2#3$}
    \setlength{\fboxsep}{2pt}        
    \sbox2{\hspace{-.5pt}
      \colorbox{#1}{\usebox0}
    }%
    \dp2=\dp0 \ht2=\ht0 \wd2=\wd0     
    \box2                             
  \endgroup                           
}

\newcommand{\highlightred}[1]{%
  \colorbox{icolor!50}{$\displaystyle#1$}}

\def\undertilde#1{\mathord{\vtop{\ialign{##\crcr
$\hfil\displaystyle{#1}\hfil$\crcr\noalign{\kern1.5pt\nointerlineskip}
$\hfil\tilde{}\hfil$\crcr\noalign{\kern1.5pt}}}}}

\usepackage{authblk}

\newcommand{\diff}{\mathrm{d}}

\usepackage{sectsty} 
\sectionfont{\color{cool_green}\selectfont}
\subsectionfont{\color{cool_green}\selectfont}
\subsubsectionfont{\color{cool_green}\selectfont}
\paragraphfont{\color{gray}\selectfont}
\date{\today}

\title{
A tutorial on the Bayesian statistical approach to inverse problems
}
\author[1]{Faaiq G. Waqar}
\author[2*]{Swati Patel}
\author[3*]{Cory M. Simon}
\affil[1]{School of Electrical Engineering and Computer Science. Oregon State University. Corvallis, OR. USA.}
\affil[2]{Department of Mathematics. Oregon State University. Corvallis, OR. USA.}
\affil[3]{School of Chemical, Biological, and Environmental Engineering. Oregon State University. Corvallis, OR. USA.}
\affil[*]{\texttt{cory.simon@oregonstate.edu}, \texttt{patelswa@oregonstate.edu}}

\begin{document}

\sloppy 

\maketitle

\begin{abstract}
Inverse problems are ubiquitous in the sciences and engineering. Two categories of inverse problems concerning a physical system are (1) estimate parameters in a model of the system from observed input-output pairs and (2) given a model of the system, reconstruct the input to it that caused some observed output. Applied inverse problems are challenging because a solution may (i) not exist, (ii) not be unique, or (iii) be sensitive to measurement noise contaminating the data.

Bayesian statistical inversion (BSI) is an approach to tackle ill-posed and/or ill-conditioned inverse problems. Advantageously, BSI provides a ``solution'' that (i) quantifies uncertainty by assigning a probability to each possible value of the unknown parameter/input and (ii) incorporates prior information and beliefs about the parameter/input.

Herein, we provide a tutorial of BSI for inverse problems, by way of illustrative examples dealing with heat transfer from ambient air to a cold lime fruit. First, we use BSI to infer a parameter in a dynamic model of the lime temperature from measurements of the lime temperature over time. Second, we use BSI to reconstruct the initial condition of the lime from a measurement of its temperature later in time. We demonstrate the incorporation of prior information, visualize the posterior distributions of the parameter/initial condition, and show posterior samples of lime temperature trajectories from the model. Our tutorial aims to reach a wide range of scientists and engineers.

\end{abstract}

\begin{center}
    \includegraphics[width=0.4\textwidth]{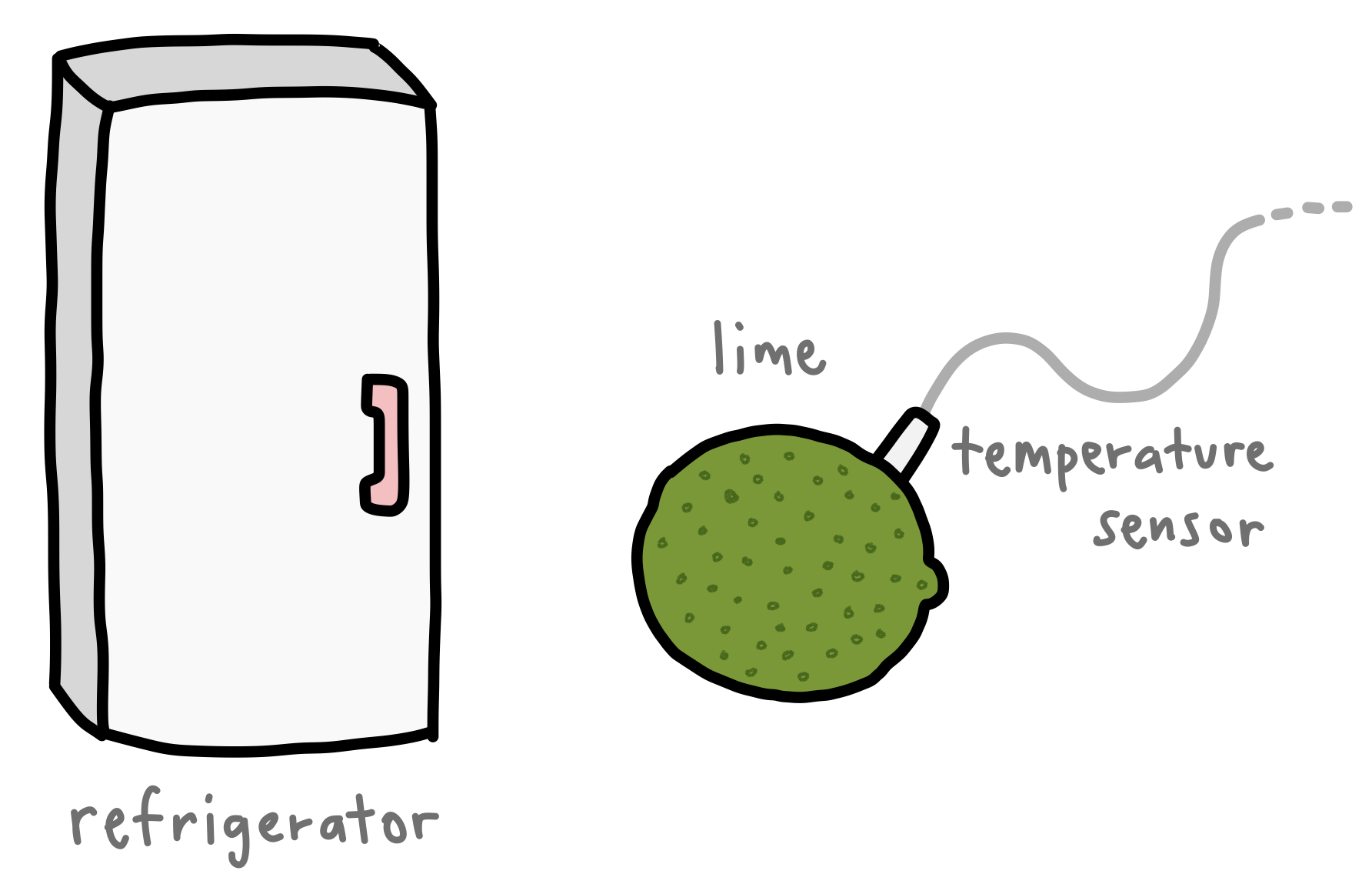}
\end{center}

\newpage 

\section{Introduction}
\subsection{Mathematical models of physical systems}
In the sciences and engineering, we often apply physical principles to develop a mathematical model of a physical system in order to capture some phenomenon of interest.
Physics-based models are useful for acquiring insights and understanding about a system, explaining observations, guiding future experiments, discovering new scientific questions, and making predictions \cite{epstein2008model}. 

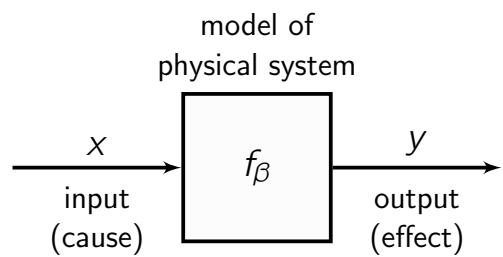
\begin{figure}[h!]
    \centering
     \begin{tikzpicture}[auto, node distance=3.25cm,>=latex']
        \node [block, label={[align=center]model of\\physical system}] (g) {\large $f_\beta$};
        \node [input, name=input, left of=g] {};
        \node [output, name=output, right of=g] {};
        
        \draw [->, line width=0.5mm] (input) -- node[above] {\large $x$} node[below, align=left] {\begin{tabular}{c} input \\ (cause) \end{tabular}} (g);
        \draw [->, line width=0.5mm] (g) -- node[above] {\large $y$} node[below] {\begin{tabular}{c} output \\ (effect) \end{tabular}} (output);
    \end{tikzpicture}
    \caption{\textbf{The mathematical model} $y=f_\beta(x)$, parameterized by $\beta$, predicts the output $y$ of a physical system in response to an input $x$.}
    \label{fig:input_output}
\end{figure}

Abstractly, a physics-based model of a system is typically an operator $f_\beta: x\mapsto y$ that (i) predicts the output $y$ in response to any given input $x$ and (ii) contains a vector $\beta$ of physical parameters characterizing the system. See Fig.~\ref{fig:input_output}. 
 

The input and output could be scalars, vectors, or functions. Evaluating the operator $f_\beta$ at an input $x$ could constitute evaluating a function, solving a system of algebraic equations, multiplying by a matrix, solving a differential equation, evaluating an integral, or running a computer simulation. \cite{aster2018parameter,groetsch1999inverse} Note, the abstraction of a physical system in Fig.~\ref{fig:input_output} does not require the system to have mass and/or energy input/output; by definition, the input is merely a causal factor set in the beginning of the experiment, and the output is an effect/result of the input \cite{iglesias2014inverse}.


\subsection{The forward problem}
The \emph{forward problem} is, given the model structure $f_\beta(x)$ and its parameters $\beta$, predict the output $y$ of the physical system in response to a given input $x$---predict the effect of a cause.

The forward problem is solved by evaluating $f_\beta(x)$.

Forward problems tend to be (but are not always\footnote{For example, in chaotic dynamical systems, the future state is highly sensitive to the initial state \cite{strogatz2018nonlinear}.}) well-posed and well-conditioned (ie., have a unique solution that is insensitive to error in the data $x$ input to the problem) \cite{kabanikhin2008definitions}. 

\subsection{Inverse problems}
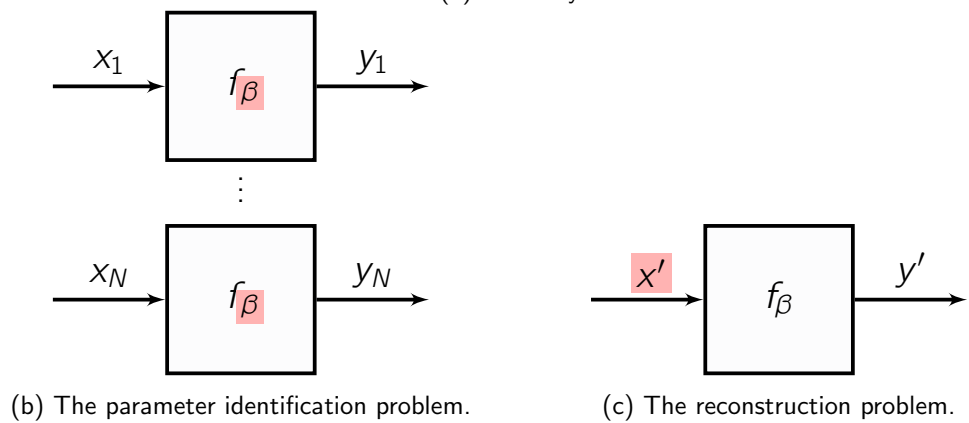
\begin{figure}[h!]
    \centering

        \begin{subfigure}[b]{\textwidth}
        \begin{tabular}{ c c c c c }
            \toprule
             inverse problem & \highlightred{\text{unknown}}
             & data & model structure, $f_\beta(x)$ & model parameter, $\beta$ \\  \midrule
             parameter identification & $\beta$ & $\{(x_i,y_{i})\}_{i=1}^N$ & {\color{gadflygreen} \bf \checkmark } & {\color{icolor} X} \\
             reconstruction & $x^\prime$ & $y^\prime$ & {\color{gadflygreen} \bf \checkmark } & {\color{gadflygreen} \bf \checkmark } \\ \bottomrule 
        \end{tabular}
        \caption{summary} \label{tab:unify_framing}
    \end{subfigure}

        \begin{subfigure}[b]{0.425\textwidth}
        \centering 
     \begin{tikzpicture}[auto, node distance=2.5cm,>=latex']
        \node [block] (g) {\large $f_{\highlight[icolor!50]{\beta}}$};
        \node [input, name=input, left of=g] {};
        \node [output, name=output, right of=g] {};
        
        \draw [->, line width=0.5mm] (input) -- node[above] {\large $x_1$} (g);
        \draw [->, line width=0.5mm] (g) -- node[above] {\large $y_1$} (output);
    \end{tikzpicture} 

    $\vdots \vspace{0.5\baselineskip}$
    
      \begin{tikzpicture}[auto, node distance=2.5cm,>=latex']
        \node [block] (g) {\large $f_{\highlight[icolor!50]{\beta}}$};
        \node [input, name=input, left of=g] {};
        \node [output, name=output, right of=g] {};
        
        \draw [->, line width=0.5mm] (input) -- node[above] {\large $x_N$} (g);
        \draw [->, line width=0.5mm] (g) -- node[above] {\large $y_N$} (output);
    \end{tikzpicture} 
    
    \caption{The parameter identification problem.}
    \end{subfigure}
    \begin{subfigure}[b]{0.425\textwidth}
        \centering 
     \begin{tikzpicture}[auto, node distance=2.5cm,>=latex']
        \node [block] (g) {\large $f_\beta$};
        \node [input, name=input, left of=g] {};
        \node [output, name=output, right of=g] {};
        
        \draw [->, line width=0.5mm] (input) -- node[above] {\large $\highlight[icolor!50]{x^\prime}$} (g);
        \draw [->, line width=0.5mm] (g) -- node[above] {\large $y^\prime$} (output);
    \end{tikzpicture} \caption{The reconstruction problem.}
    \end{subfigure}
     
    \caption{\textbf{Two categories of inverse problems}. }
    \label{fig:inv_prob_overview}
\end{figure}

\epigraph{Mathematicians find it difficult to define ``inverse problem'', yet most recognize one when they see it.}{\textit{Charles W. Groetsch} \cite{groetsch1999inverse}}
Two categories of \emph{inverse problems} \cite{groetsch1999inverse,aster2018parameter,neto2012introduction,fernandez2013bayes,ito2014inverse,denisov1999elements,mueller2012linear,tenorio2017introduction} are (see Fig.~\ref{fig:inv_prob_overview}) the:
\begin{enumerate}
    \item \emph{parameter identification} problem: determine the parameters $\beta$ characterizing a system that produced a set of observed input-output pairs $\{(x_i, y_{i})\}_{i=1}^N$.
    \item \emph{reconstruction problem}: determine/reconstruct the input $x^\prime$ that produced an observed system output $y^\prime$, given the model 
    parameters $\beta$. Ie., predict the cause of an observed effect. 
\end{enumerate}

Inverse problems are ubiquitous in the sciences and engineering. 
Parameter identification is often involved in the development of a quantitatively predictive model.
Reconstruction problems emerge when measurements we ideally wish to make in order to probe a system are infeasible, inaccessible, invasive, dangerous, and/or destructive; eg.\ 
(i) sensing or imaging using indirect measurements\footnote{An elementary example is exploiting a thermal expansion model of mercury to infer the temperature of the air from a measurement of the volume of mercury in a thermometer \cite{kaipio2006statistical,heilio2016mathematical}.},
(ii) making inferences about the interior of a domain from measurements on its boundary, and
(iii) time reversal: reconstructing the past state of a system from a measurement of its current state. \cite{groetsch1999inverse,aster2018parameter}

A famous inverse problem, one of reconstruction, is ``can one hear the shape of a drum?'' \cite{kac1966can,protter1987can,gordon1992one,gordon1996you}.

\begin{mybox}[ 
floatplacement=t,
breakable,arc=0mm,
colback=cool_gray!10,
colframe=cool_green, 
fonttitle=\bfseries,
before upper={\parindent15pt}
]{Can one hear the shape of a drum?}
\small 
\noindent Suppose we strike a drumhead with a stick at $t=0$.
The force induces transverse waves in the drumhead, which transmit to the air to produce longitudinal waves in the air (sound). \\

\noindent Treating the drumhead as a homogeneous, elastic membrane, the wave equation is a mathematical model for the vertical displacement $u(x, t)$ of the membrane at a point $x \in \Omega$ and time $t \geq 0$:
\begin{align}
    & \frac{\partial^2 u}{\partial t^2} = c \Delta_{x} u, \quad x \in\Omega, t\geq 0& \text{\color{gray} wave equation} \label{eq:wave} \\
    & u\rvert_{\partial \Omega}=0. & \text{\color{gray} clamped boundary conditions} \nonumber 
\end{align}
The parameter $c>0$ is the ratio of the spatially-uniform tension in the membrane to its density; $\Delta_{x}$ is the Laplace operator; $\Omega \subset \mathbb{R}^2$ defines the geometry of the membrane ($\partial \Omega \subset \mathbb{R}^2$ is the boundary of $\Omega$). Damping is neglected. Eqn.~\ref{eq:wave} is subject to an initial position and velocity set by the striking drumstick. \cite{kreyszig2009advanced} \\

\noindent The pure tones the drumhead can produce are dictated by standing wave solutions of the wave equation, $e^{i\omega t} \phi(x)$. Substituting into eqn~\ref{eq:wave}, we find the frequency $\omega$ of a pure tone of the drumhead is related to an eigenvalue of the Laplace operator on the domain $\Omega$:
\begin{align}
    & c \Delta_{x} \phi = -\omega^2 \phi, \quad x \in \Omega & \text{\color{gray} an eigenvalue problem}   \label{eq:eigenvalue_prob} \\
    & \phi\rvert_{\partial \Omega}=0 . & \text{\color{gray} clamped boundary conditions}  \nonumber
\end{align}
According to eqn.~\ref{eq:eigenvalue_prob}, the membrane, characterized by its physical property $c$ and its geometry $\Omega$, can produce a discrete spectrum of vibration frequencies $\omega_1 \geq \omega_2 \geq \cdots$. \cite{kac1966can} \\

\noindent The conventional \emph{forward problem} is: given the physical property $c$ and geometry $\Omega$ of the membrane, use eqn.~\ref{eq:eigenvalue_prob} to determine the spectrum of vibration frequencies $\omega_1 \geq \omega_2 \geq \cdots$ it can produce. In undergraduate mathematics courses, students solve the forward problem analytically for a rectangular or circular drum head. \cite{kuttler1984eigenvalues} \\

\noindent An \emph{inverse problem} is: given the physical property $c$ of the membrane and its spectrum of vibration frequencies $\omega_1 \geq \omega_2 \geq \cdots$, use eqn.~\ref{eq:eigenvalue_prob} to determine its geometry $\Omega$. 
The set of pure tones a drum can produce contains information about the geometry of its membrane; but, it is not obvious if these pure tones (and $c$) \emph{uniquely determine} the geometry of the membrane. This inverse problem, ``can one hear the shape of a drum?'' was popularized by Kac in 1966, but it was not until 1992 that Gordon, Webb, and Wolpert found two non-congruent domains with an identical spectrum of vibration frequencies \cite{gordon1996you}, shown below.

\begin{minipage}[t]{\linewidth}
    \centering 
    \includegraphics[width=0.5\textwidth]{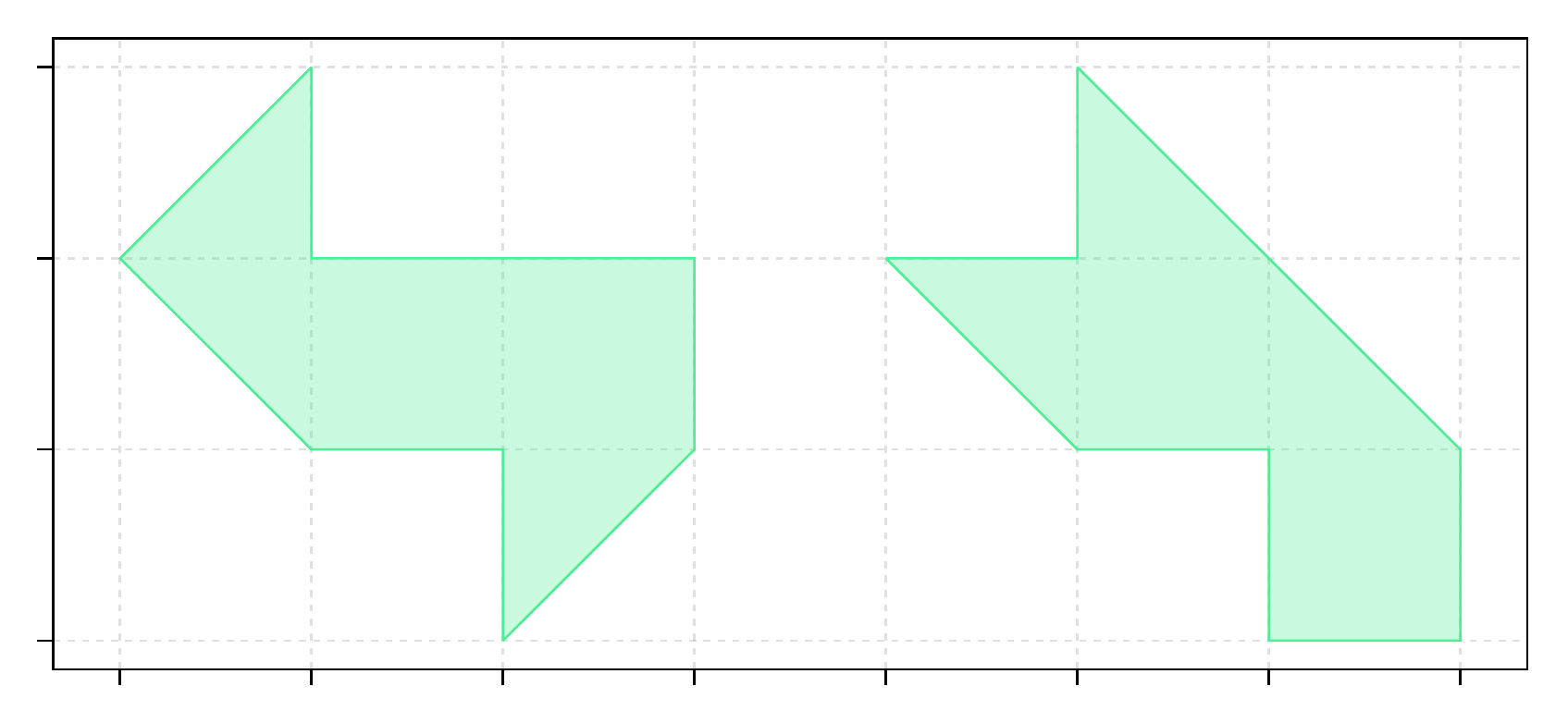}
\end{minipage}
    
ie., one cannot hear the shape of a drum \cite{gordon1996you}.
\end{mybox}

Classically, a solution to an inverse problem is sought by adjusting the input $x$ or parameter $\beta$ until the model output(s) match(es)/fit(s) the observed system output(s) eg., via least squares \cite{chavent2010nonlinear,lines1984review}, to achieve consistency between the model and the data \cite{kennedy2001bayesian}. 


\subsubsection{Challenges in applied inverse problems}
As opposed to theoretical, \emph{exact} inverse problems, in \emph{applied} inverse problems \cite{sabatier2000past}, we must cope with noise contaminating the measurements of the system output.
This noise may originate from: 
(i) an imperfect measurement instrument and/or 
(ii) variance in the system output in response to a repeated input due to (a) inherent stochasticity and/or (b) unrecognized or poorly-controlled, and thus unaccounted-for in the model, inputs/conditions that influence the output.
More, the model $f_\beta(x)$ may inadequately approximate physical reality---perhaps, in part, owing to (ii)b. \cite{kennedy2001bayesian}

Applied inverse problems are particularly interesting and challenging because, in contrast to forward problems, they are often ill-posed and/or ill-conditioned  \cite{kabanikhin2008definitions,maclaren2019can}; a solution to an applied inverse problem:
\begin{itemize}
    \item may not exist, owing to noise in the measured output and model inadequacy. 
    \begin{itemize}
        \item in reconstruction problems, the observed output may fall outside the image of $f_\beta$, so that no physically-viable input is consistent with it.
        \item in parameter identification, a parameter $\beta$ giving a model $f_\beta(x)$ that reproduces all observed input-output pairs may not exist.
    \end{itemize}
    \item may not be unique. 
    \begin{itemize}
        \item in reconstruction problems, many different inputs may be consistent with the observed output, owing to a many-to-one $f_\beta$. 
        \item in parameter identification, (i) the number of observed input-output pairs may be insufficient to fully constrain the parameters or (ii) the model may be inherently unidentifiable (ie.\ no amount of data can fully constrain the parameters) \cite{guillaume2019introductory,maclaren2019can,chis2011structural}. 
    \end{itemize}
    \item may depend discontinuously on or be sensitive to measurement noise contaminating the data. 
    Even if the solution to the inverse problem exists and is unique, noise contaminating measurements of the system output propagates onto and corrupts the solution to the inverse problem. 
    In ill-conditioned inverse problems \cite{heilio2016mathematical}, small realizations of noise in the data produce large changes in the solution.
\end{itemize}

To tackle applied inverse problems, these challenges implore us to account for the noise contaminating measurements of the output, reframe our concept of a ``solution'', and quantify our uncertainty in the solution \cite{dashti2017bayesian,tarantola1982inverse}.

\subsection{Bayesian statistical inversion (BSI)}
Bayesian statistical inversion (BSI) \cite{kaipio2006statistical,stuart2010inverse,tarantola2005inverse,tarantola1982inverse,idier2013bayesian,ulrych2001bayes,fitzpatrick1991bayesian} is a versatile framework to tackle [possibly] ill-posed and/or ill-conditioned applied inverse problems. BSI has two key advantages. 
First, BSI allows us to incorporate prior (ie.\ before data is collected) information and/or beliefs about the input/parameter into our solution to the inverse problem. Second, BSI yields a probabilistic solution to inverse problems that quantifies uncertainty about the input/parameter.

\bookemoji \emph{key references for BSI}: \cite{kaipio2006statistical,calvetti2018inverse}.

\paragraph{Modeling uncertainty.} To model uncertainty about the unknown input/parameter, we treat it as a random variable and model its probability distribution. In this Bayesian view, the probability assigned to each region of input/parameter space reflects our degree of belief that the input/parameter falls in that region. This belief is based on some combination of subjective belief and objective information. \cite{trotta2008bayes,ghosh2006introduction} 
Loosely speaking, the spread (concentration) of the probability density over the input/parameter space reflects our uncertainty (certainty) about its value.

\paragraph{Prior vs.\ posterior distributions.} Our uncertainty about the input/parameter is different before and after we conduct an experiment, collect data, and compare this data with our model. 
Consequently, the input/parameter has a \emph{prior} density before the data are considered, then an updated, \emph{posterior} density after we are enlightened by the data. 

\paragraph{Modeling the data-generating process.}
To acknowledge the unobservable noise that contaminates our measurements,
we view the outcomes of our measurements of the system output (ie., the data) as realizations of random variables. 
A model of the stochastic data-generating process follows from combining (i) the mathematical model $f_\beta(x)$ of the system and (ii) a model of the probability distribution of the noise.

\paragraph{The two ingredients of BSI.}
The two key ingredients for tackling an inverse problem with BSI are:
\begin{enumerate}
    \item the \emph{prior distribution} of the input/parameter, expressing our beliefs about it before the data are collected/observed. Constructing a prior is context-dependent and involves a degree of subjectivity. A prior may be roughly categorized as diffuse, weakly informative, or informative \cite{van2021bayesian}, based on the amount of uncertainty it admits about the input/parameter. 
    An informative prior, eg.\ a Gaussian distribution with a small variance, expresses a high degree of certainty about the input/parameter. On the other hand, a diffuse prior, eg.\ a uniform distribution, spreads its density widely over input/parameter space to express a very high degree of uncertainty. \cite{van2021bayesian}
    An informative prior influences the posterior more than a diffuse prior, which ``lets the data speak for itself'' \cite{downey2021think}. Generally, as we gather more data, the influence of the prior on the posterior tends to (but does not for non-identifiable systems \cite{guillaume2019introductory}) weaken/diminish as the data overrides/overwhelms the prior.
    \item the \emph{likelihood function} of the input/parameter, giving the probability density of the data conditioned on each value of the input/parameter. The likelihood function is constructed from (i) the data and (ii) our model of the data-generating process, which constitutes (a) the forward model and (b) a probability distribution to model the noise corrupting the measurements. The likelihood quantifies the support that the data lend to each value of the unknown input/parameter. \cite{van2021bayesian}
\end{enumerate}
From these two ingredients, we next ``turn the Bayesian crank'' \cite{murphy2023probabilistic} to obtain the posterior distribution of the unknown input/parameter.

\paragraph{The BSI solution to an inverse problem.}
The BSI ``solution'' to the inverse problem, the \emph{posterior distribution} of the unknown input/parameter, follows from the prior density and likelihood function of the input/parameter via Bayes' theorem. 
The posterior density of the unknown input/parameter gives the probability that the input/parameter falls in any given region of input/parameter space, conditioned on the data.
The posterior updates the prior in light of the data, offers a compromise between the prior and likelihood, and constitutes the raw solution to the inverse problem that quantifies uncertainty about the input/parameter through its spread. \cite{dashti2017bayesian} 
In practice, some inference engine (eg. Markov Chain Monte Carlo sampling) is usually employed to approximate the posterior \cite{murphy2023probabilistic}.
We may summarize the posterior distribution of the input/parameter with a \emph{credible region} (eg., a highest-density region \cite{hyndman1996computing}) that contains some large fraction $\alpha$ of the posterior density and thus credibly contains (given the assumptions in the likelihood and prior hold) the input/parameter (precisely, with probability $\alpha$) \cite{koch2007introduction}. 
Note, a Bayesian credible interval for the parameter/input is distinct from and arguably more intuitive and natural than a frequentist confidence interval \cite{hespanhol2019understanding}.

\subsection{Our contribution: a tutorial of BSI, illustrated on inverse problems of heat transfer}
We provide a tutorial of BSI to solve inverse problems of both parameter identification and reconstruction while incorporating prior information and quantifying uncertainty. Our tutorial is by way of examples regarding a simple, intuitive physical process: convective heat transfer from ambient air to a cold lime fruit. 
We aim to engage a wide range of scientists and engineers. 

In Sec.~\ref{sec:setup}, we describe our experimental setup: (i) we take a cold lime fruit out of a refrigerator and allow it to exchange heat with indoor air via natural convection; (ii) a probe inserted into the lime measures its internal temperature.
In Sec.~\ref{sec:model}, we pose a mathematical model governing the time-evolution of the lime temperature, based on Newton's ``law'' of cooling. The model contains a single parameter. 
To describe the data-generating process, in Sec.~\ref{sec:dgp}, we augment this model with a probabilistic model of making a noisy measurement of the lime temperature with the imperfect temperature probe.
Next, in Sec.~\ref{sec:param_id}, we employ BSI to infer the parameter in the dynamic model of the lime temperature (an overdetermined inverse problem), using time series data of the lime temperature.
We then employ the model of the lime temperature with the inferred parameter to tackle two time reversal problems: (Sec.~\ref{sec:time_reversal_a}, ill-conditioned) reconstruct the initial temperature of the lime, given a measurement of its current temperature and the time it has been outside of the refrigerator and   (Sec.~\ref{sec:time_reversal_b}, underdetermined) reconstruct the initial temperature of the lime and the duration it has been out of the refrigerator, given a measurement of its current temperature. 
The solution to each inverse problem under BSI expresses uncertainty through the posterior probability density of the parameter/initial condition. We assess the fidelity of the solution to the reconstruction problems by comparing with the measured initial condition held out from the BSI procedure. 
From these concrete, instructive examples, we intend for readers to recognize inverse problems in their research domain where BSI may be employed to incorporate prior information and quantify uncertainty.

\section{A tutorial of Bayesian statistical inversion (BSI)}
As a tutorial of BSI to solve inverse problems while incorporating prior information and quantifying uncertainty, we tackle a variety of inverse problems pertaining to heat exchange between ambient air and a lime fruit via natural convection.


\subsection{Experimental setup: heat exchange between a lime and the air}
\label{sec:setup}
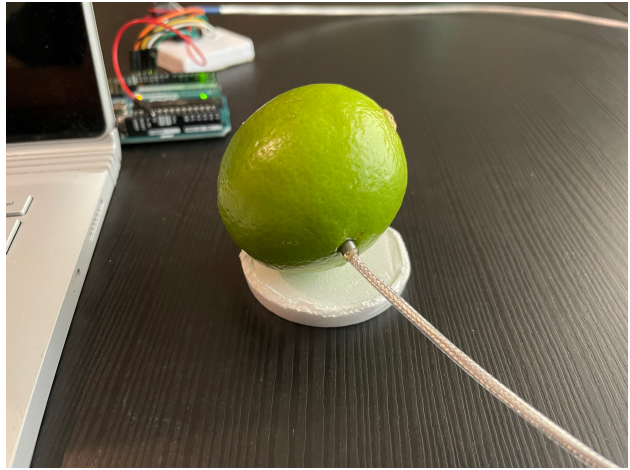
\begin{figure}[h!]
    \centering
    \begin{tikzpicture}
    \node [mybox] (box){%
        \begin{minipage}{0.6\textwidth}
          \textbf{the system}: a lime fruit
    
          \textbf{the process}: heat transfer via natural convection
    
          \textbf{the measurement instrument}: a temperature probe

          \textbf{a data point}: $(t, \theta_{\text{obs}})$
        \end{minipage}
    };
    \end{tikzpicture}%

    \includegraphics[width=0.5\textwidth]{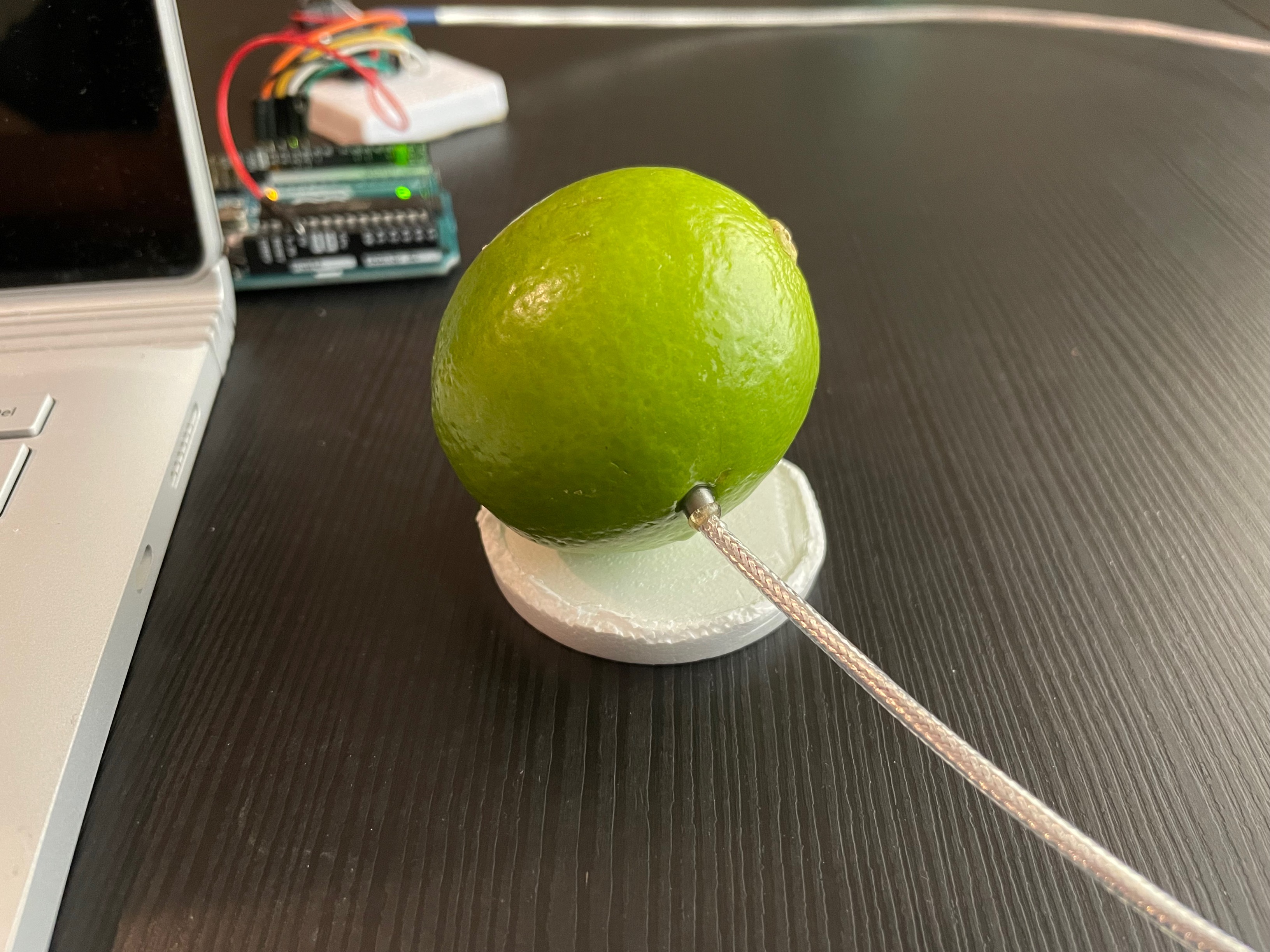}

    \caption{\textbf{Experimental setup.} An initially-cold lime fruit rests on a styrofoam slab and exchanges heat with the indoor air via natural convection. A temperature sensor inserted into the lime can measure its internal temperature at any given time $t$, producing a data point $(t, \theta_{\text{obs}})$. 
    }
    \label{fig:exptl_setup}
\end{figure}

We allowed a lime fruit ($\sim$5\,cm diameter) to reside in a refrigerator for several days. 
Then, at time $t:=t_0$ [min], we removed the lime from the refrigerator, placed it on a thin slab of insulating styrofoam, and allowed it to exchange heat with the indoor air via natural convection. 
A resistance-based temperature sensor inserted into the lime\footnote{
To avoid the early-time data reflecting the [short] time dynamics of the temperature probe coming into thermal equilibrium with the lime, owing to its nonzero thermal mass, we inserted the probe into the lime \emph{before} we placed it in the refrigerator, so it begins thermally equilibrated with the lime.} allows us to measure its internal temperature at any given time $t$ to generate a data point $(t, \theta_{\text{obs}})$ (``obs'' = observed). Fig.~\ref{fig:exptl_setup} shows our experimental setup and Sec.~\ref{sec:temp_sensing} explains our temperature sensor and Arduino microcontroller setup.

\subsection{The mathematical model of the lime temperature}
\label{sec:model}
We develop a mathematical model governing $\theta=\theta(t)$ [$^\circ$C], the temperature of the lime as a function of time $t \geq t_0$ [min], as it exchanges heat with the bath of ambient air at bulk (ie., far from the lime) temperature $\theta^{\text{air}}$ [$^\circ$C].
The mechanism of lime-air heat transfer is natural convection, which is the combined effect \cite{cengel2012ebook} of both (i) heat conduction and (ii) heat transport via motion of the air adjacent to the lime, driven by buoyant forces arising from changes in the density of the air due to changes in its temperature \cite{bird2002transport}. 
The initial temperature of the lime is $\theta(t_0)=:\theta_0$.

\paragraph{Assumptions.}
We make several simplifying assumptions:
\begin{itemize}
    \item the temperature of the lime is spatially uniform\footnote{We estimate the Biot number \cite{cengel2012ebook} to be $\text{Bi}:=hr/k\approx 0.6$, based on: (i) our measurement of the radius of the lime $r\approx 2.5$\,cm, (ii) the reported thermal conductivity of a lime $k \approx 0.595$\,J/(s$\cdot$m$\cdot^\circ$C) \cite{ikegwu2009thermal}, and (iii) an estimated heat transfer coefficient for natural convection via a gas, $h\approx15$\,J/(s$\cdot$m$^2\cdot ^\circ$C) \cite{KOSKY2013259,cengel2012ebook}. 
    }
    \item the bulk temperature of the air (a bath) $\theta^{\text{air}}$ is constant (sufficiently far from the lime)
    \item heat conduction between the lime and the styrofoam surface on which it sits is negligible
    \item the mass of the lime is constant (eg., a negligible loss of moisture over time)
    \item the heat capacity of the lime is constant with temperature
    \item heat released/consumed due to chemical reactions (eg., oxidation) is negligible
    \item the temperature probe inserted into the lime has a negligible thermal mass
    \item the rate of heat exchange between the air and the lime is governed by Newton's ``law'' of cooling
\end{itemize}

\paragraph{Newton's ``law'' of cooling.}
Newton's ``law'' of cooling \cite{vollmer2009newton,rees1988cooling,o1990newton,bohren1991comment} prescribes the rate of heat transfer [J/min] from the air to the lime at time $t\geq t_0$ as proportional to the difference in temperature between the lime and the air (the thermodynamic driving force for heat transfer) at that time:
\begin{equation}
    \text{rate of heat transfer from air to lime} \propto \theta^{\text{air}}-\theta(t).
\end{equation}
The precise rate, $hA[\theta^{\text{air}}-\theta(t)]$, depends on two (assumed, constant) parameters: (i) the surface area of the lime in contact with the air, $A$ [cm$^2$] and (ii) the [natural] convective heat transfer coefficient, $h$ [J/($^\circ$C$\cdot$min$\cdot$cm$^2$)], between the air and the surface of the lime \cite{cengel2012ebook}.

\paragraph{Differential equation model.}
Conservation of energy applied to the lime gives its change in temperature $\diff \theta=\diff \theta (t)$ over a differential change in time $\diff t$:
\begin{equation}
    C\diff \theta(t) = hA[\theta^{\text{air}}-\theta(t)]\diff t \quad \text{for } t\geq t_0, \label{eq:model_differential_form}
\end{equation}
with $C$ [J/$^\circ$C] the thermal mass of the lime.
Eqn.~\ref{eq:model_differential_form} balances the amount [J] of sensible heat stored in the lime (left) and amount [J] of heat transferred to the lime (right) over the time interval $[t, t+dt)$.
Thus, a first-order differential equation describes the time-evolution of the lime temperature \cite{vollmer2009newton,rees1988cooling}:
\begin{align}
    & \lambda \frac{\diff \theta}{\diff t}=\theta^{\text{air}}- \theta(t) \quad \text{for } t\geq t_0 \label{eq:model} \\
    &\theta (t=t_0)=\theta_0. \nonumber
\end{align}
The single, lumped parameter $\lambda:=C/(hA)$ [min] of the model is a time constant governing the dynamics of heat transfer between the lime and the air.

\paragraph{Analytical solution to the model.}
Eqn.~\ref{eq:model} admits an analytical solution through a variable transformation, then integration:
\begin{equation}
    \theta (t)=\theta^{\text{air}}+(\theta_0-\theta^{\text{air}})e^{-(t-t_0)/\lambda}, \quad \text{for } t\geq t_0. \label{eq:model_soln}
\end{equation} 
Starting at $\theta_0$, the lime temperature $\theta(t)$ monotonically approaches the air temperature ($\lim_{t\rightarrow\infty}\theta(t)=\theta^{\text{air}}$) as the temperature difference between the lime and air decays exponentially.
The parameter $\lambda$ is a time scale for the lime to reach thermal equilibrium with the air. Specifically, the lime temperature reaches a fraction $e^{-1}\approx 0.63$ of the way to the air temperature after a duration $t-t_0=\lambda$ out of the refrigerator. 

\begin{figure}[h!]
    \centering
    \includegraphics[width=0.5\textwidth]{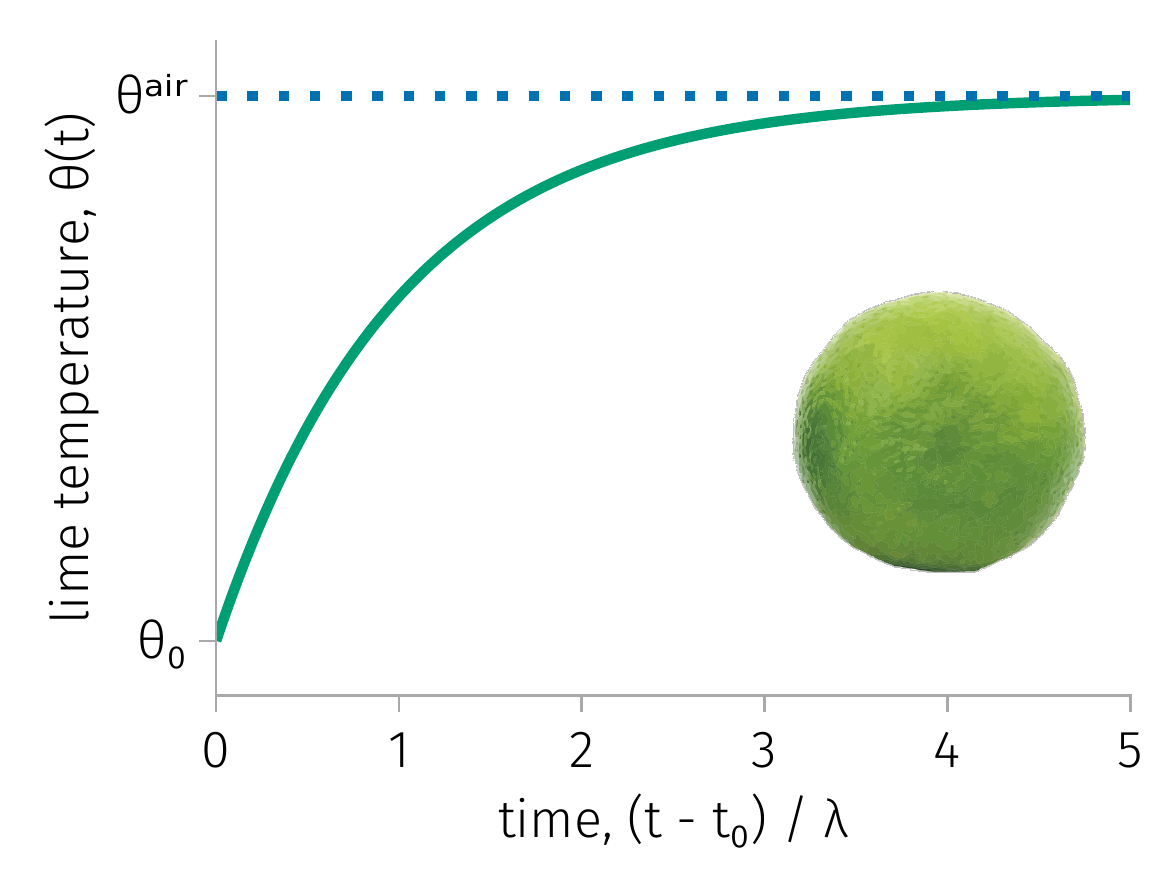}
    \caption{\textbf{The solution to the mathematical model of the lime temperature}, $\theta(t; \lambda, t_0, \theta_0, \theta^{\text{air}})$.}
    \label{fig:model_soln}
\end{figure}

Fig.~\ref{fig:model_soln} shows the solution to the model of the lime temperature, which we write as $\theta(t; \lambda, t_0, \theta_0, \theta^{\text{air}})$ to emphasize its dependence on the parameter $\lambda$, the initial condition $(t_0, \theta_0)$, and the air temperature $\theta^{\text{air}}$.

\subsection{The probabilistic model of the data-generating process}
\label{sec:dgp}
Consider the process of employing our temperature probe, an imperfect measurement instrument, to measure the lime temperature at time $t$, giving a data point $(t, \theta_{\text{obs}})$.

We treat the measured temperature $\theta_{\text{obs}}$ as a realization of a random variable $\Theta_{\text{obs}}$ owing to two sources of stochasticity. 
First, \emph{measurement noise}: unobservable noise originating from the temperature probe corrupts the measurement. 
Second, \emph{residual variability}: under repeated experiments with identical conditions $(t_0, \theta_0, \theta^{\text{air}})$, the lime temperature at time $t$ may exhibit variance due to variable conditions/inputs that are poorly-controlled or unrecognized, and thus unaccounted for in the model $\theta(t)$ \cite{kennedy2001bayesian}. For example, (i) the air temperature $\theta^{\text{air}}$ is not perfectly controlled and may fluctuate over time and (ii) the opening and closing of doors in the building may introduce fluctuating air currents in the room, making the heat transfer coefficient $h$ change over time. 

We model the noise in the observed lime temperature as a random variable $E_\sigma$ additive to the model prediction, independent among repeated measurements, and having an identical distribution over time.
Then, our probabilistic model of the data-generating process is:
\begin{equation}
    \Theta_{\text{obs}} = \theta(t; \lambda, t_0, \theta_0, \theta^{\text{air}}) + \Epsilon_\sigma, \label{eq:Theta_obs}
\end{equation}
where $E_\sigma$ is a zero-centered Gaussian with variance $\sigma^2$:
\begin{equation}
    \Epsilon_\sigma \sim \mathcal{N}(0, \sigma^2). \label{eq:epsilon_sigma}
\end{equation}
Eqn.~\ref{eq:Theta_obs} assumes the time scale for the temperature probe to thermally equilibrate with the lime is negligibly small, avoiding a time delay.

The random variable $E_\sigma$ can capture noise emanating from both the measurement instrument and residual variability. However, eqn.~\ref{eq:Theta_obs} assumes the mean measured lime temperature at time $t$ over repeated experiments with the same conditions $(t_0, \theta_0, \theta^{\text{air}})$ is given by the model $\theta(t; \lambda, t_0, \theta_0, \theta^{\text{air}})$. Ie., our data-generating model neglects the possibility of \emph{model discrepancy}, a nonzero difference between (a) the expected measured lime temperature at time $t$ over multiple experiments with the same conditions $(t_0, \theta_0, \theta^{\text{air}})$ and (b) the model prediction of the lime temperature, $\theta(t; \lambda, t_0, \theta_0, \theta^{\text{air}})$. \cite{kennedy2001bayesian} 

Under eqn.~\ref{eq:Theta_obs}, the probability density function governing the distribution of the measured lime temperature $\Theta_{\text{obs}}$ at time $t$, given $\lambda, t_0, \theta_0, \theta^{\text{air}}, \sigma$, is obtained by translating the density of the noise in eqn.~\ref{eq:epsilon_sigma} to center it at the model prediction:
\begin{equation}
     \pi_{\text{likelihood}}( \theta_{\text{obs}} \mid \lambda, t_0, \theta_0, \theta^{\text{air}}, \sigma) = \frac{1}{ \sigma \sqrt{2\pi} } \exp\left[ 
     -\frac{1}{2}\left(\frac{\theta_{\text{obs}}-\theta(t ; \lambda, t_0, \theta_0, \theta^{\text{air}})}{ \sigma}\right)^2
     \right]
     .\label{eq:likelihood_single_data}
\end{equation}

\subsection{Inverse problem I: parameter identification}
\label{sec:param_id}

\begin{mybox}[ 
floatplacement=t,
breakable,arc=0mm,
colback=cool_gray!10,
colframe=cool_green, 
fonttitle=\bfseries,
before upper={\parindent15pt}
]{Overview of problem and approach}
\small 
\noindent 
\textbf{Task}: employ BSI to infer the parameter $\Lambda$ (ie., $\lambda$ treated as a random variable) characterizing the lime, appearing in the model of the lime temperature in eqn.~\ref{eq:model_soln}, using data from a heat transfer experiment. \\

First, we estimate $\lambda$ with a back-of-the-envelope calculation and use this estimate to construct a weakly informative prior distribution of $\Lambda$. \\

To admit our uncertainty about the variance of the measurement noise in our model of the data-generating process in eqn.~\ref{eq:Theta_obs}, we treat $\Sigma^2$ as an unknown parameter to also be inferred from the data. We impose a diffuse prior distribution on $\Sigma^2$, granted support based on the noise characteristics of the temperature probe. \\

Next, we conduct a heat transfer experiment and collect data (see Sec.~\ref{sec:setup}). Our measurements defining the conditions of the experiment are:
(i) the initial temperature of the lime, giving data $(t_0=0, \theta_{0, \text{obs}})$ and 
(ii) the air temperature, giving data $\theta^{\text{air}}_\text{obs}$. 
To admit these are noisy measurements, we treat the initial temperature of the lime $\Theta_0$ and air temperature $\Theta^{\text{air}}$ as random variables to be inferred from the data, but impose informative prior distributions on them based on these measurements.
Over the course of the experiment, we measure the lime temperature, giving time series data \thedata that provide information about $\Lambda$. \\

Finally, we use Bayes' theorem to construct the posterior distribution of $(\Lambda, \Theta_0, \Theta^{\text{air}}, \Sigma)$ in light of the data, sample from it using a Markov Chain Monte Carlo algorithm, and obtain a credible interval for the parameter $\Lambda$ that quantifies posterior uncertainty about its value. \\

\textbf{Quick overview}:
\begin{itemize}
    \item \emph{data}: the measured initial temperature of the lime $(t_0=0, \theta_{0, \text{obs}})$, air temperature $\theta^{\text{air}}_\text{obs}$, and lime temperature over the course of the experiment \thedata
    \item \emph{random variables to infer from the data:} the parameter $\Lambda$, the initial lime temperature $\Theta_0$, the air temperature $\Theta^{\text{air}}$, the variance of the measurement noise $\Sigma^2$
    \item \emph{sources of priors}: 
    $\Lambda$: back-of-the-envelope calculation; 
    $\Theta_0, \Theta^{\text{air}}$: our noisy measurements of them; 
    $\Sigma^2$: precision of temperature sensor
\end{itemize}

\textbf{Summary of results:} See Fig.~\ref{fig:param_id}.

\end{mybox}

Classically, this inverse problem is overdetermined because a parameter $\lambda$ giving a model $\theta(t; \lambda, t_0, \theta_0, \theta^{\text{air}})$ that passes through all data points \thedata does not exist\footnote{Note, we do not attempt to determine each $C$, $h$, and $A$ in eqn~\ref{eq:model_differential_form} from the data \thedata but rather the lumped parameter $\lambda:=C/(hA)$. The former would make the system unidentifiable because any $(C, h, A)$ that give the same $\lambda$ produce the same model $\theta(t)$.}.

\subsubsection{The prior distributions}
Before the data \thedata are considered, in BSI, we must express our prior beliefs and information about the value of each variable via a prior probability density function.

\paragraph{The parameter, $\Lambda$.} A back-of-the-envelope estimate of $\lambda = C/(hA)$ is $\approx 1$\,hr, based on the following. The diameter of the lime, approximated as a sphere, is $\sim$5\,cm. The mass of the lime is $\sim$100\,g. 
The specific heat of a lime is reported as $\sim$4.0 kJ/[kg$\cdot ^\circ$C] \cite{ikegwu2009thermal,mukama2020thermophysical}. 
A typical heat transfer coefficient $h$ for natural convection via gas is 15\,J/[s$\cdot$m$^2\cdot ^\circ$C] \cite{KOSKY2013259,cengel2012ebook}.  

We specify a weakly informative prior density $\pi_{\text{prior}}(\lambda)$ as that of a Gaussian distribution (i) centered at our back-of-the-envelope estimate of $\lambda$, (ii) with a high variance to reflect our low confidence in this rough estimate, and (iii) truncated below zero to enforce positivity:
\begin{equation}
    \Lambda \sim \mathcal{N}_{> 0} \left(1 \text{ hr}, (0.3\text{ hr})^2 \right). \label{eq:lambda_prior}
\end{equation}
See Fig.~\ref{fig:param_id_prior_posterior}.

\paragraph{The experimental conditions, $\Theta_0$ and $\Theta^{\text{air}}$.}
We impose informative prior distributions on the initial lime temperature $\Theta_0$ and air temperature $\Theta^{\text{air}}$ based on our (noisy) measurements of them:
\begin{align}
    \Theta_0 & \sim \mathcal{N}(\theta_{0, \text{obs}}, \sigma^2) \\ 
    \Theta^{\text{air}} & \sim \mathcal{N}(\theta_{\text{obs}}^{\text{air}}, \sigma^2).
\end{align}

\paragraph{The variance of the measurement noise, $\Sigma^2$.}
Our prior distribution of the standard deviation of the measurement noise, reflecting our beliefs about the precision of the temperature probe, is
\begin{equation}
    \Sigma \sim \mathcal{U}([0\,^\circ \text{C}, 1\,^\circ \text{C}]),
\end{equation} where $\mathcal{U}(\cdot)$ is a uniform distribution over the interval $\cdot$.

\paragraph{The joint prior distribution.} The joint prior distribution of all of the unknowns for this inverse problem factorizes since we imposed independent priors, corresponding to plausible assumptions, incl.\ that the parameter $\lambda$ of the lime has no causality link with the air temperature.
\begin{equation}
    \pi_{\text{prior}}(\lambda, \theta_0, \theta^{\text{air}}, \sigma)=
    \pi_{\text{prior}}(\lambda)
    \pi_{\text{prior}}(\theta_0)
    \pi_{\text{prior}}(\theta^{\text{air}})
    \pi_{\text{prior}}(\sigma).
\end{equation}

\lightbulbemoji The prior distribution $\pi_{\text{prior}}(\lambda, \theta_0, \theta^{\text{air}}, \sigma)$ summarizes the information and beliefs we have about the unknowns $(\lambda, \theta_0, \theta^{\text{air}}, \sigma)$ at this stage, before the data \thedata are considered. 

\subsubsection{The data and likelihood function}
\begin{figure}
    \centering
    \begin{subfigure}{0.48\textwidth}
        \includegraphics[width=\textwidth]{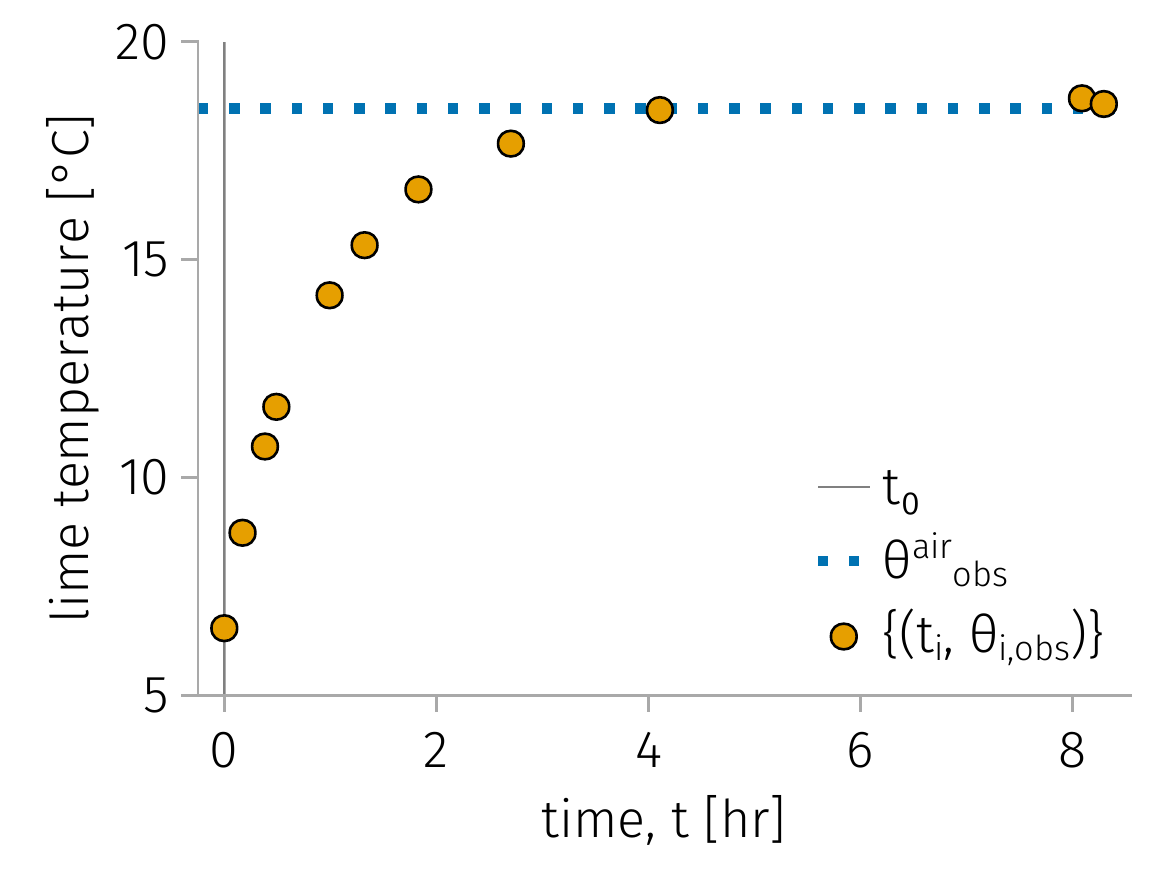} \caption{} \label{fig:param_id_data}
    \end{subfigure}

    \begin{subfigure}{0.48\textwidth}
        \includegraphics[width=\textwidth]{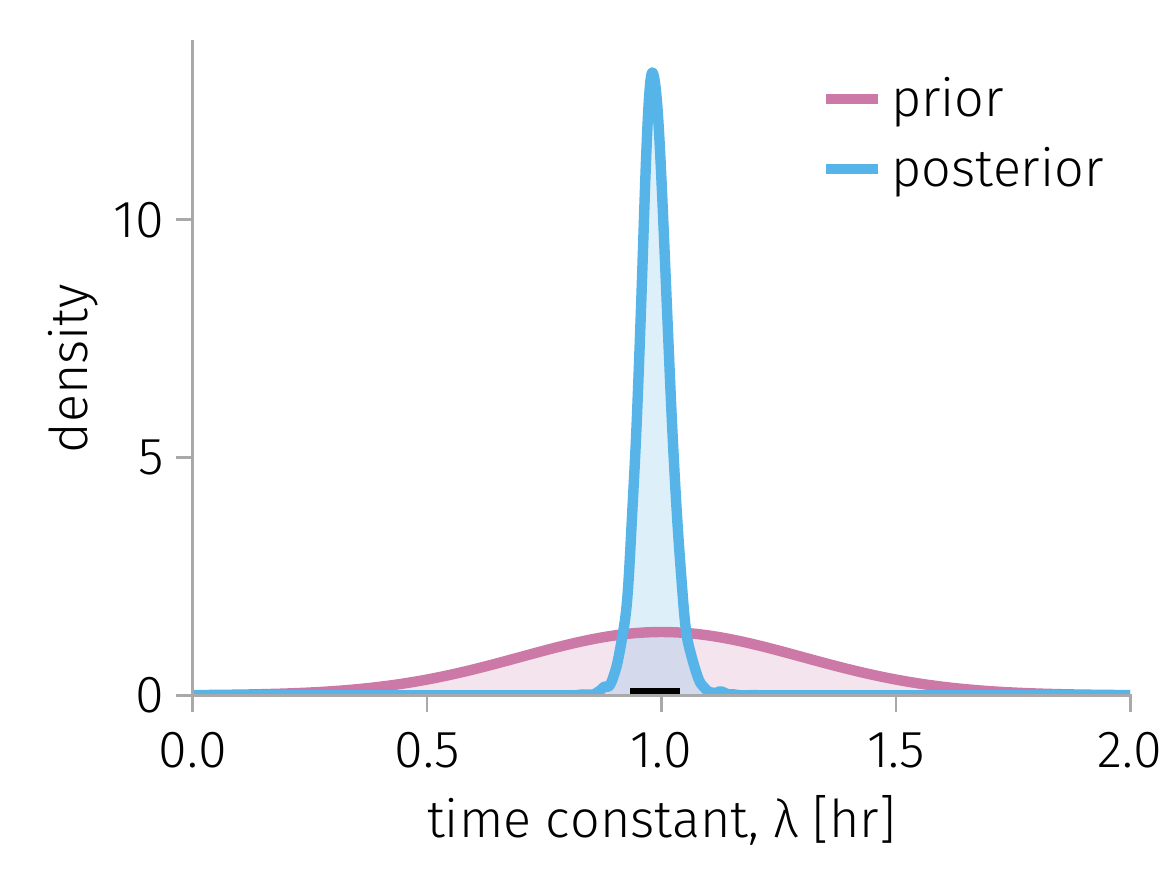} \caption{} \label{fig:param_id_prior_posterior}
    \end{subfigure}
    \begin{subfigure}{0.48\textwidth}
        \includegraphics[width=\textwidth]{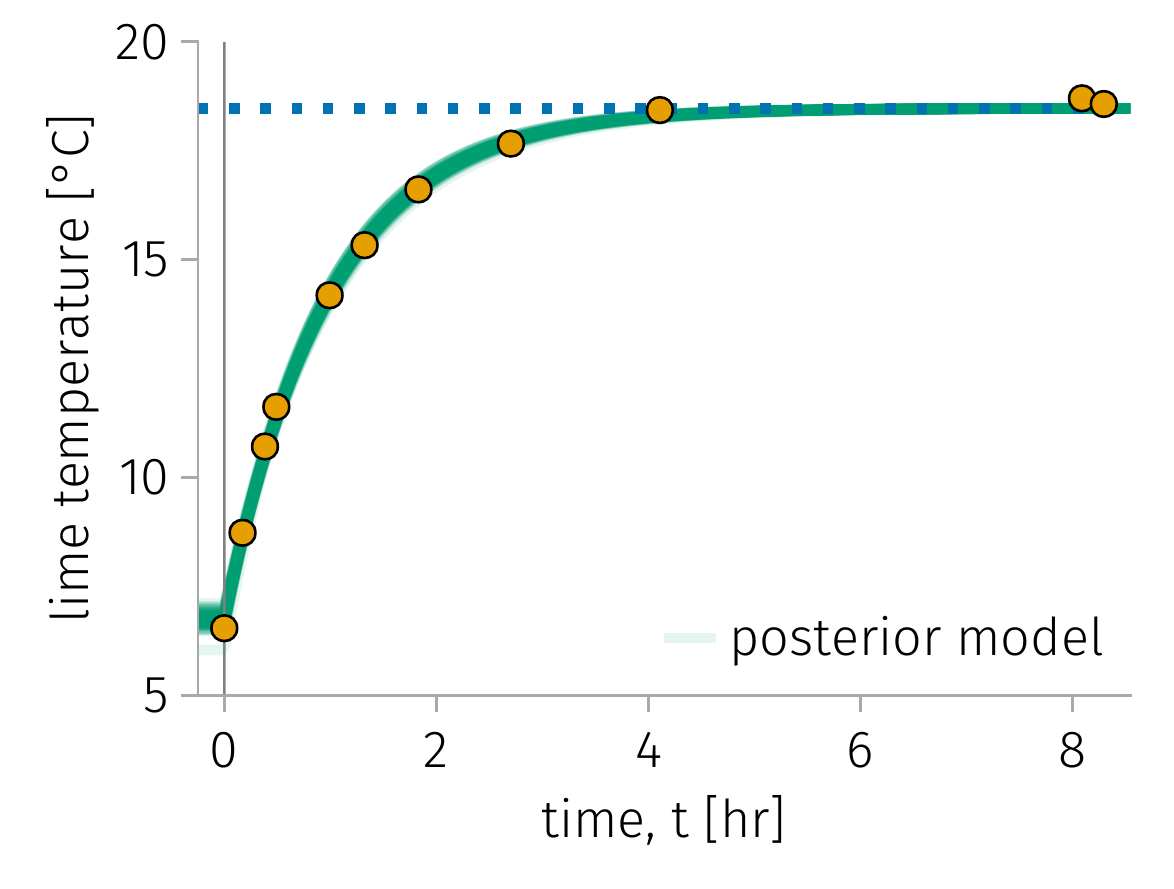} \caption{} \label{fig:param_id_trajectory}
    \end{subfigure}
    \caption{\textbf{Inverse problem I: parameter identification}: infer the parameter $\Lambda$ in the model of the lime temperature, $\theta(t; \lambda, t_0, \theta_0, \theta^{\text{air}})$, given measurements of the conditions of the experiment, $(t_0=0, \theta_{0, \text{obs}})$ and $\theta^{\text{air}}_{\text{obs}}$, and time series data of the lime temperature \thedata. 
    (a) The data.
    (b) The prior and posterior distributions of $\Lambda$. The black bar shows the equal-tailed, 90\% posterior credible interval for $\Lambda$.
    (c) A sample of 100 model trajectories $\theta(t; \lambda, t_0=0, \theta_0, \theta^{\text{air}})$, with $(\lambda, \theta_0, \theta^{\text{air}})$ sampled  from the posterior.
    }
    \label{fig:param_id}
\end{figure}

\paragraph{The data.} The data from the heat transfer experiment are displayed in Fig.~\ref{fig:param_id_data}:
\begin{itemize}
    \item the initial condition of the lime $(t_0=0, \theta_{0, \text{obs}})$
    \item the air temperature $\theta^{\text{air}}_\text{obs}$
    \item the lime temperature over the course of the experiment \thedata.
\end{itemize}

\paragraph{The likelihood function.}
The likelihood function gives the probability density of the data \thedata conditioned on each possible value of the parameters $\Lambda=\lambda$ and $\Sigma=\sigma$, and experimental conditions $\Theta_0=\theta_0$ and $\Theta^{\text{air}}=\theta^{\text{air}}$.
We construct the likelihood from the (i) data \thedata and (ii) model of the data-generating process in eqn.~\ref{eq:likelihood_single_data}. 
The likelihood function is:
\begin{align}
   \pi_{\text{likelihood}}(\theta_{1, \text{obs}}, ..., \theta_{N, \text{obs}} \mid  \lambda, \theta_0, \theta^{\text{air}}, \sigma) \propto
        \displaystyle \prod_{i=1}^{N} 
        \exp \left[-\frac{1}{2}\left(\frac{\theta_{i, \text{obs}}-\theta(t_i ; \lambda, t_0=0, \theta_0, \theta^{\text{air}})}{\sigma}\right)^2 \right].  
    \label{eq:likelihood_param_id}
\end{align}
The likelihood factorizes because we model the measurement noise $E_\sigma$ as an independent random variable. Note, inherently, the likelihood is conditioned on the model structure as well. 

\lightbulbemoji Since we possess the data \thedata at this stage, we (i) view the likelihood as a function of the unknowns $(\lambda, \theta_0, \theta^{\text{air}}, \sigma)$ and (ii) interpret it as a measure of the support the data \thedata lend to each value of the unknowns, $(\lambda, \theta_0, \theta^{\text{air}}, \sigma)$. \cite{van2021bayesian}

\subsubsection{The posterior distribution}
The (joint) \emph{posterior density} governs the probability distribution of the unknowns $(\Lambda, \Theta_0, \Theta^{\text{air}}, \Sigma)$ conditioned on the time series data \thedata. 
By Bayes' theorem \cite{koch2007introduction}, the posterior density is proportional to the product of the likelihood function and prior density:
\begin{multline}
    \pi_{\text{posterior}}(\lambda, \theta_0, \theta^{\text{air}}, \sigma \mid \theta_{1, \text{obs}}, ..., \theta_{N, \text{obs}}) \propto \\ 
    \pi_{\text{likelihood}}(\theta_{1, \text{obs}}, ..., \theta_{N, \text{obs}} \mid \lambda, \theta_0, \theta^{\text{air}}, \sigma)\pi_{\text{prior}}(\lambda, \theta_0, \theta^{\text{air}}, \sigma).
   \label{eq:posterior_param_id}
\end{multline}
Note, because we will employ a Markov chain Monte Carlo algorithm to sample from the posterior and approximate it empirically, we do not need to know the normalizing factor; these samplers only require \emph{ratios} of posterior densities. 

We are particularly interested in the posterior distribution of the parameter $\Lambda$, with $(\Theta_0, \Theta^{\text{air}}, \Sigma)$ marginalized out.

\lightbulbemoji The posterior density of the unknowns $\pi_{\text{posterior}}(\lambda, \theta_0, \theta^{\text{air}}, \sigma \mid \theta_{1, \text{obs}}, ..., \theta_{N, \text{obs}})$:
\begin{itemize}
    \item is the raw solution to this parameter identification problem, as it assigns a probability to each region of $(\lambda, \theta_0, \theta^{\text{air}}, \sigma)$-space to reflect our posterior degree of belief that the unknowns $(\lambda, \theta_0, \theta^{\text{air}}, \sigma)$ fall in that region.
    \item represents an \emph{update} to our prior density $\pi_{\text{prior}}(\lambda, \theta_0, \theta^{\text{air}}, \sigma)$, in light of the data \thedata.
    \item reflects a compromise between (i) our prior knowledge and beliefs about $(\lambda, \theta_0, \theta^{\text{air}}, \sigma)$ and (ii) the support the data \thedata lend to $(\lambda, \theta_0, \theta^{\text{air}}, \sigma)$, according to our model of the data-generating process.
\end{itemize}

\paragraph{Remark on sources of posterior uncertainty about $\Lambda$.}
Uncertainty about $\Lambda$ remains even after the data \thedata are considered. 
Sources of this posterior uncertainty owe to a lack of data (small $N$) coupled with two causes of noise captured through our model of the data-generating process in eqn.~\ref{eq:Theta_obs} \cite{kennedy2001bayesian}:
(i) measurements of the lime temperature being corrupted by [unobservable] noise from using an imperfect temperature sensor and
(ii) unrecognized and/or poorly-controlled inputs/conditions that influence the lime temperature and vary \emph{over the course of the experiment}. 
However, our posterior distribution does \emph{not} capture uncertainty due to 
(i) residual variability: unrecognized and/or poorly-controlled inputs/conditions that influence the lime temperature and vary \emph{from experiment-to-experiment} (because we conduct only a single experiment) or 
(ii) model inadequacy: when $\theta(t)$ does not faithfully predict the expected (over many experiments under the same conditions $(t_0, \theta_0, \theta^{\text{air}})$) measured lime temperature (perhaps, in part owing to (i)), which would introduce bias.

\paragraph{Sampling from the posterior.} 
We employ a \emph{Markov Chain Monte Carlo} (MCMC) algorithm, the No-U-Turn Sampler (NUTS) \cite{hoffman2014no} implemented in \texttt{Turing.jl} \cite{ge2018t} in Julia \cite{bezanson2012julia}, to obtain samples from the joint posterior distribution in eqn.~\ref{eq:posterior_param_id} in order to (i) approximate it with an empirical distribution using kernel density estimation and (ii) compute means and credible intervals of the unknowns.
Over four independent chains, we collect 2\,500 samples/chain, with the first half discarded for burn-in.

\begin{mybox}[ 
floatplacement=t,
breakable,arc=0mm,
colback=cool_gray!10,
colframe=black!90, 
fonttitle=\bfseries,
before upper={\parindent15pt}
]{Markov chain Monte Carlo (MCMC) sampling from the posterior distribution}
\small 
\noindent
In a general setting for parameter inference via BSI, let
\begin{itemize}
    \item $U \in \mathbb{R}^K$ be the random vector of the $K$ unknown parameters in the inverse problem
    \item $d \in \mathbb{R}^N$ be the data vector of noisy measurements/observations.
\end{itemize}
The posterior density of $U$ follows from Bayes' theorem \cite{koch2007introduction}:
\begin{equation}
    \pi(u) := \pi_{\text{posterior}}(u \mid d) = \frac{\pi_{\text{likelihood}}(d \mid u) \pi_{\text{prior}}(u)}{\pi_{\text{evidence}}(d)}.
\end{equation}
The denominator, \emph{the evidence}, is the probability density of the data---a constant that serves as a normalizing factor for the numerator:
\begin{equation}
    \pi_{\text{evidence}}(d) = \int \pi_{\text{likelihood}}(d \mid u) \pi_{\text{prior}}(u) du.
\end{equation}
Usually, we cannot analytically evaluate this $K$-dimensional integral. 
If $K$ is large, it may be intractable to use numerical cubature to approximate $\pi_{\text{evidence}}(d)$ as well. 
The same difficulty may arise for the integral to (i) compute the mean of the posterior or (ii) marginalize out a subset of the unknowns we are less concerned with. \cite{murphy2023probabilistic}
 \\

MCMC methods permit us to obtain samples $u_1, ..., u_n$ from the posterior density $\pi(u)$ while only knowing a function to which it is \emph{proportional} (as in eqn.~\ref{eq:posterior_param_id}). This circumvents the need to compute $\pi_{\text{evidence}}(d)$. 
From the samples $u_1, ..., u_n$, we can
(i) construct an empirical posterior distribution using kernel density estimation \cite{chen2017tutorial} and 
(ii) approximate (a) the mean of the posterior from the mean of the samples and (b) an equal-tailed credible interval of any unknown from the percentiles of its samples.
Note, we can construct the empirical \emph{marginal} posterior distribution, of a subset of unknowns, trivially by ignoring the remaining dimensions of the sampled vectors. \\

The idea behind an MCMC method is to (i) 
construct a Markov chain $U_1,U_2,...$ whose
(a) state space is the parameter space, and
(b) transition kernel governing the probability of transitioning from one state $u$ to another $u^\prime$ endows the chain with a stationary distribution equal to the posterior distribution $\pi(u)$ then 
(ii) simulate the Markov chain to obtain a realization $u_1, ..., u_n$, regarded as (serially-correlated) samples from $\pi(u)$. \cite{murphy2023probabilistic,sherlock2010random,van2021bayesian,roy2020convergence}
 \\

Perhaps the simplest MCMC simulation algorithm to understand is random walk Metropolis \cite{murphy2023probabilistic,sherlock2010random}. 
Here, a realization $u_1, ..., u_n$ of a Markov chain $U_1, ..., U_n$ is obtained by iterating a stochastic process of ``propose then accept-or-reject'' $n$ times. 
Suppose $u$ is the current state in the chain. 
We \emph{propose} to move to a new state $u^\prime$, chosen randomly according to an isotropic random walk starting at $u$. 
Ie., the proposed new state is a random variable $U^\prime=u+\Delta U$ where $\Delta U \sim \mathcal{N}(0, \sigma^2 I)$.
We \emph{accept} this proposed state transition with probability
\begin{equation}
    \min \left[ 1, \frac{\pi(u^\prime)} {\pi(u)} \right] \label{eq:acceptance}
\end{equation} and \emph{reject} it otherwise, staying at $u$. 
This rule (i) always accepts proposed moves ``uphill'' to a state $u^\prime$ with higher density than $u$ and (ii) occasionally accepts moves ``downhill''.
Note, the rule only requires the ratio of the densities of the two states. Hence, the normalization factor cancels and $\pi_{\text{evidence}}(u)$ is not needed.
Together, this proposal distribution and acceptance rule specifies a transition kernel $\pi(u^\prime  \mid u)$ that grants the Markov chain $U_1, U_2, ...$ a stationary density equal to $\pi(u)$. Consequently, $u_1, ..., u_n$ are serially-correlated samples from the posterior $\pi(u)$. \cite{murphy2023probabilistic}
The scale parameter $\sigma$ in the random walk proposal distribution dictates the efficiency of the sampling, in terms of the amount of serial correlation in the samples. If $\sigma$ is too small, too many proposed random walk steps are required to explore the state space. If $\sigma$ is too large, too many proposals will be to visit regions with low density, which will be rejected, making the walker stay in place. Both extremes make the sampler inefficient. \cite{sherlock2010random,roberts2001optimal} \\

NUTS \cite{hoffman2014no}, an extension of Hamiltonian Monte Carlo (HMC) \cite{betancourt2017conceptual}, is a more efficient MCMC sampler than random walk Metropolis owing to its more intelligent proposal scheme than a random walk in state space. 
\end{mybox}

\subsubsection{Summary of results}
Fig.~\ref{fig:param_id_data} shows all data from the heat transfer experiment that we employ to infer the parameter $\Lambda$ with BSI.

\rocketemoji Fig.~\ref{fig:param_id_prior_posterior} compares (i) the prior distribution of the parameter $\Lambda$ with (ii) its updated, (marginal) empirical posterior distribution constructed via kernel density estimation. The bar shows the \emph{equal-tailed 90\% posterior credible interval} for $\Lambda$, $[0.94\,\text{hr}, 1.01\,\text{hr}]$.
By definition, the true parameter $\lambda$ of the lime is situated in this interval with 90\% probability, falls below it with 5\% probability, and falls above it with 5\% probability. The width of the interval, then, reflects our posterior uncertainty about $\Lambda$.
This interpretation is predicated upon our model of the data-generating process and prior assumptions holding.

Fig.~\ref{fig:param_id_trajectory} illustrates the posterior distribution over functions $\theta(t)$ modeling the lime temperature by showing a random sample of 100 realizations of models for the lime temperature, $\theta(t;\lambda, t_0=0, \theta_0, \theta^{\text{air}})$, with $(\lambda, \theta_0, \theta^{\text{air}})$ a sample from the posterior distribution. The models fit the data \thedata well and exhibit little variance (see the residual plot in Fig.~\ref{fig:residuals}; the mean posterior model of the lime temperature matches each data point to within $\pm$0.25$^\circ$C), reflecting low uncertainty about the parameter $\lambda$ in light of the data. 


We found little correlation (Pearson correlation 0.01) between $\Lambda$ and $\Sigma$ in the joint posterior distribution. The (marginal) posterior distributions of $\Lambda$ and $\Sigma$ are well-approximated as independent Gaussian distributions $\mathcal{N}(0.98\text{ hr}, (0.02\text{ hr})^2)$ and $\mathcal{N}(0.16^\circ\text{C}, (0.03^\circ\text{C})^2)$.

To assess the efficiency of an MCMC sampler and give confidence the MCMC samples reliably approximate the posterior distribution \cite{roy2020convergence}, we drew trace plots and visualized the empirical distribution over four independent chains. See Fig.~\ref{fig:convergence_diagnostics}.

\subsection{Inverse problem IIa: time reversal}
\label{sec:time_reversal_a}

\begin{figure}[h!]
    \centering
    \centering
    \begin{subfigure}{0.48\textwidth}
        \includegraphics[width=\textwidth]{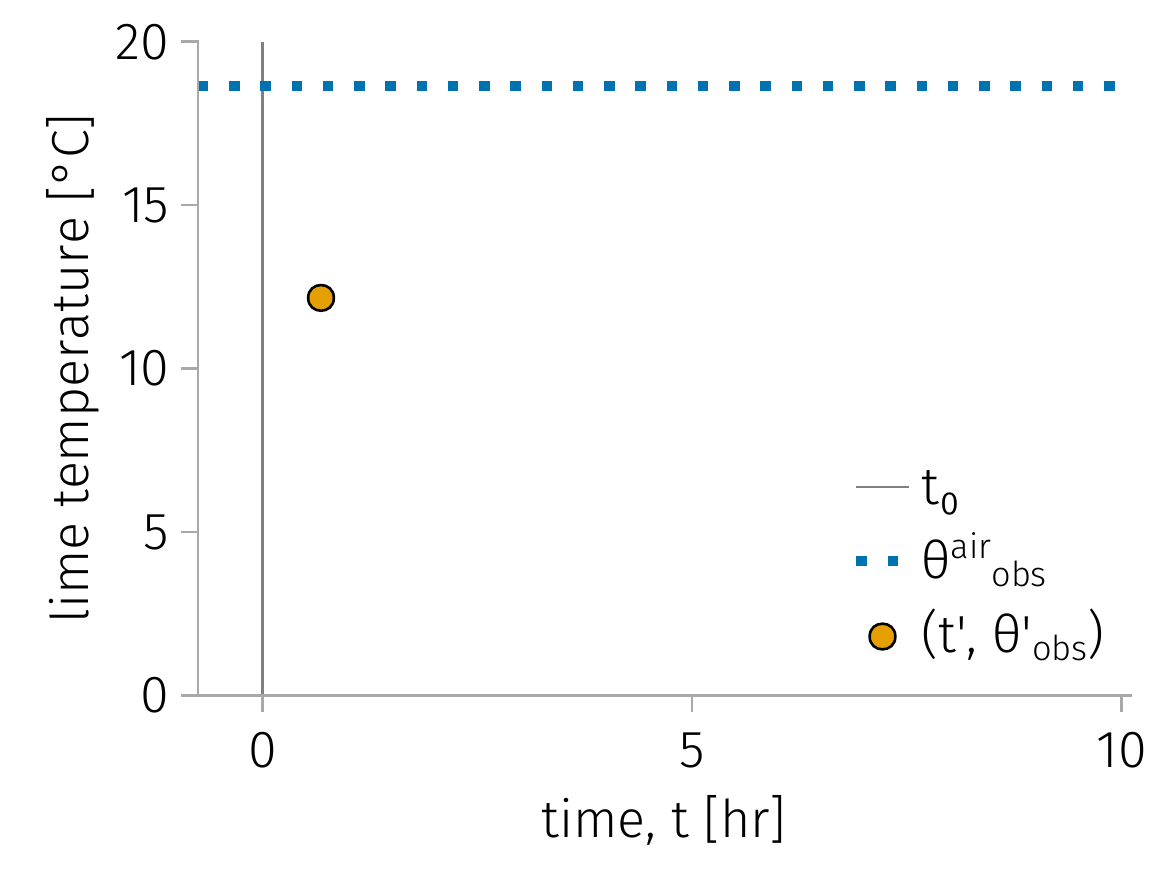} \caption{} \label{fig:tr_data}
    \end{subfigure}

    \begin{subfigure}{0.48\textwidth}
        \includegraphics[width=\textwidth]{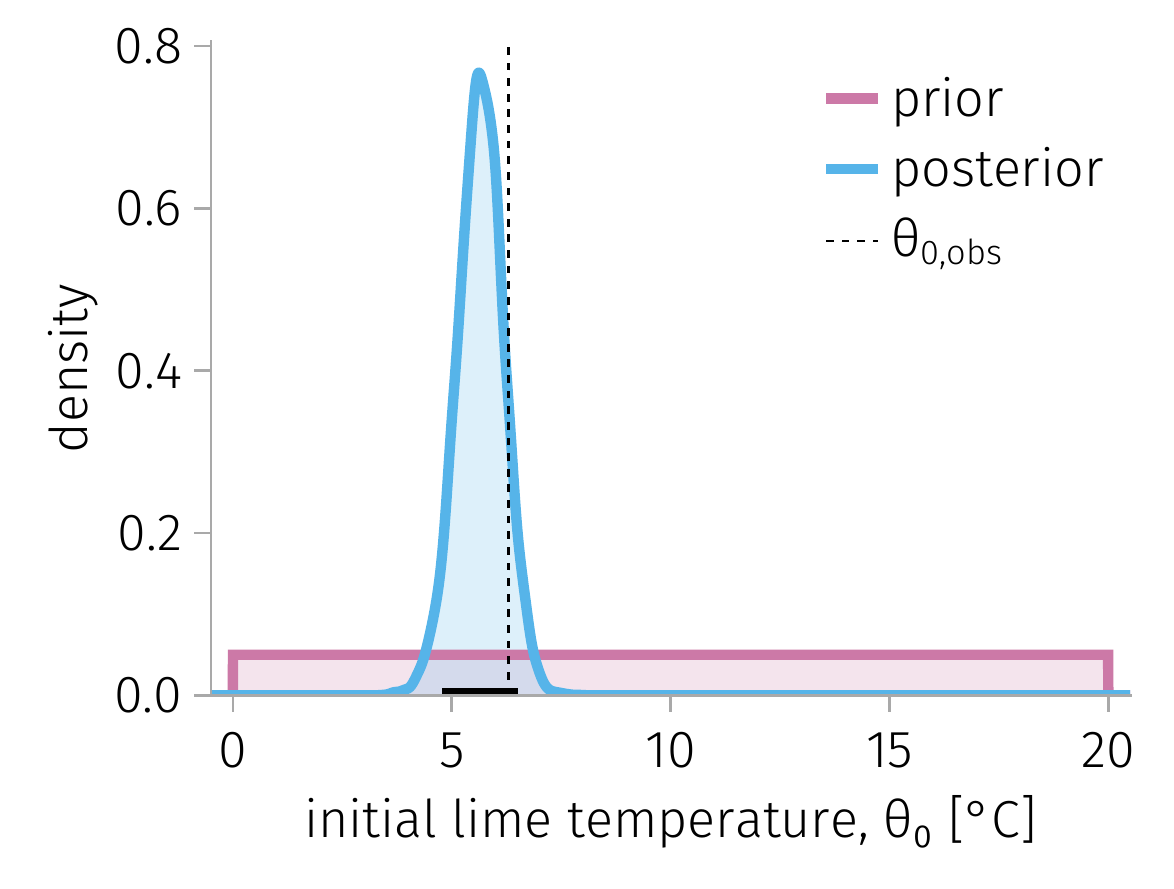} \caption{} \label{fig:tr_prior_posterior}
    \end{subfigure}
    \begin{subfigure}{0.48\textwidth}
        \includegraphics[width=\textwidth]{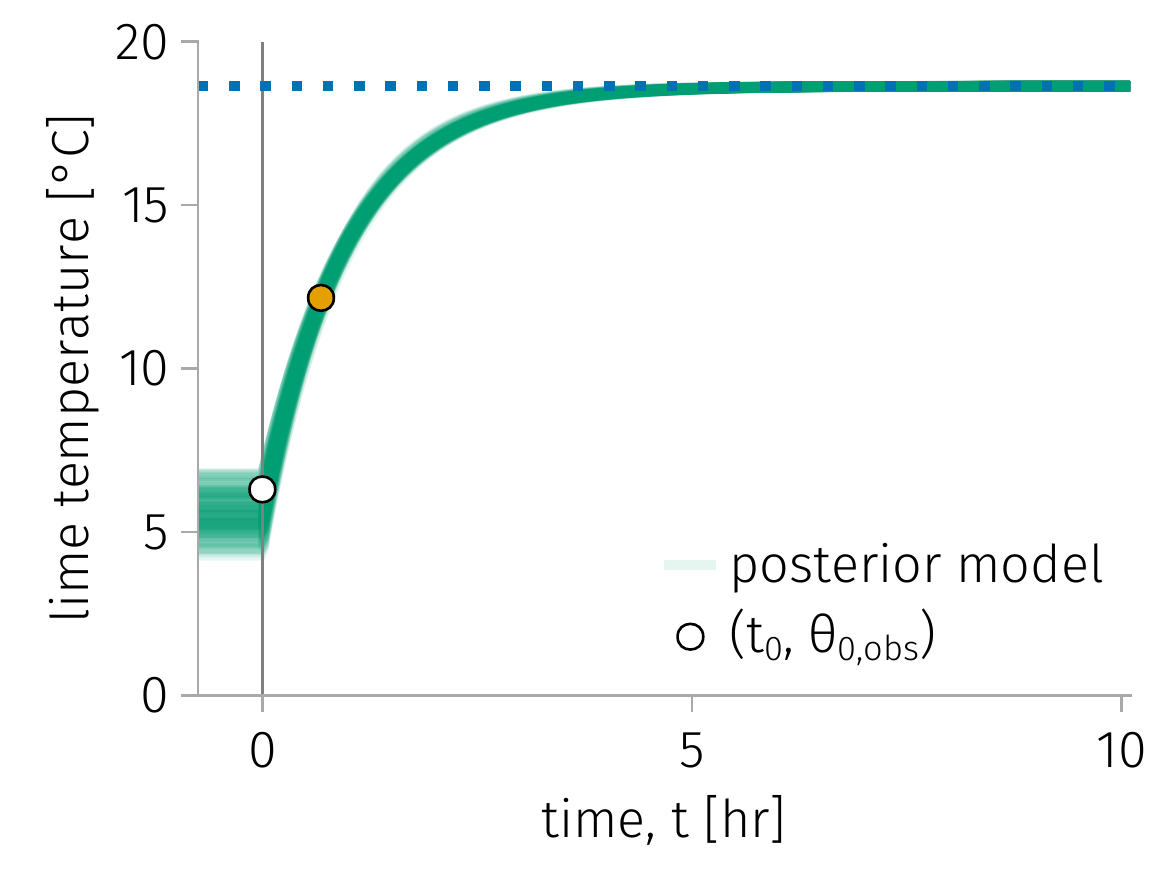} \caption{} \label{fig:tr_trajectory}
    \end{subfigure}

    \caption{\textbf{Inverse problem IIa: time reversal}: infer the initial (at time $t_0=0$) temperature $\Theta_0$ of the lime from a measurement of its temperature later, $(t^\prime, \theta_{\text{obs}}^\prime)$.
    (a) The data.
    (b) The prior and posterior distributions of $\Theta_0$. The black bar shows its equal-tailed 90\% credible interval. The vertical line shows the held-out measurement $\theta_{0, \text{obs}}$.
    (c) A sample of 100 model trajectories $\theta(t;\lambda, t_0=0, \theta_0, \theta^{\text{air}})$ from the posterior.
    } \label{fig:tr}
\end{figure}

\begin{mybox}[ 
floatplacement=t,
breakable,arc=0mm,
colback=cool_gray!10,
colframe=cool_green, 
fonttitle=\bfseries,
before upper={\parindent15pt}
]{Overview of problem and approach}
\small 
\noindent 
\textbf{Task}: employ BSI to infer the initial ($t_0=0$) temperature of the lime $\Theta_0$ (ie., $\theta_0$ treated as a random variable) based on a measurement of its temperature at time $t^\prime>t_0$, $\theta_{\text{obs}}^\prime$, and the measured air temperature $\theta_{\text{obs}}^{\text{air}}$. \\

First, we impose a diffuse prior distribution on $\Theta_0$ based on the range of temperatures encountered in refrigerators. \\

Second, we impose informative prior distributions on the parameter $\Lambda$ characterizing the lime and the variance of the measurement noise $\Sigma^2$, based on the posterior distributions from our parameter identification activity in Sec.~\ref{sec:param_id}.
``Yesterday's posterior is today's prior.'' \cite{calvetti2010subjective} \\

Next, we conduct another heat transfer experiment on the same lime (see Sec.~\ref{sec:setup}). Our measurement defining the condition of the experiment is the measured air temperature, $\theta^{\text{air}}_\text{obs}$. To (indirectly) gather information about $\Theta_0$, we measure the temperature of the lime at time $t^\prime$, ie.\ a duration $t^\prime$ after it was taken out of the refrigerator, giving $\theta^\prime_{\text{obs}}$. \\

Finally, we use Bayes' theorem to construct the posterior distribution of $(\Theta_0, \Lambda, \Theta^{\text{air}}, \Sigma)$ in light of the data, sample from it using an MCMC algorithm, and obtain a credible interval for the initial lime temperature $\Theta_0$ that quantifies posterior uncertainty about its value. \\

Note, we also measured the initial temperature of the lime, but we hold the data $(t_0=0, \theta_{0, \text{obs}})$ out from the BSI procedure to test the fidelity of the posterior distribution of $\Theta_0$. \\ 

\textbf{Quick overview}:
\begin{itemize}
    \item \emph{data}: the measured air temperature $\theta^{\text{air}}_\text{obs}$ and lime temperature $(t^\prime, \theta_{\text{obs}}^\prime)$ with $t^\prime>t_0=0$.
    \item \emph{random variables to infer from the data:} the initial lime temperature $\Theta_0$, the parameter $\Lambda$, the air temperature $\Theta^{\text{air}}$, the variance of the measurement noise $\Sigma^2$
    \item \emph{sources of priors}: 
    $\Theta_0$: range of temperatures encountered in refrigerators;
    $\Lambda, \Sigma^2$: the posterior from our parameter identification activity in Sec.~\ref{sec:param_id}; 
    $\Theta^{\text{air}}$: our noisy measurement of the air temperature.
\end{itemize}

\textbf{Summary of results:} See Figs.~\ref{fig:tr} and \ref{fig:tr_ridge}.
\end{mybox}

Classically, this is a determined time-reversal problem that becomes ill-conditioned for large $t^\prime$. To see the ill-conditioning, let $\delta \theta^\prime$ be the error in our measurement of $\theta(t^\prime)$ and $\delta \theta_0$ be the resulting error in our prediction of the initial temperature, $\hat{\theta_0}=:\theta_0+\delta\theta_0$. The errors $\delta\theta_0$ and $\delta \theta^\prime$ are related through a perturbed version of eqn.~\ref{eq:model}:
\begin{equation}
    \theta^\prime + \delta \theta^\prime =\theta^{\text{air}}+(\theta_0 + \delta \theta_0 -\theta^{\text{air}})e^{-(t-t_0)/\lambda}, \label{eq:curve}
\end{equation} implying the error in the predicted initial temperature $\delta \theta_0 = \delta \theta^\prime e^{(t-t_0)/\lambda}$ grows exponentially with the time $t^\prime>t_0$ at which we take the measurement of the lime temperature. 
This ill-conditioning is apparent from the graphical solution to this time reversal problem, of tracing the model trajectory of the lime temperature backwards in time, starting at the point $(t^\prime, \theta(t^\prime) + \delta \theta^\prime)$, back to $(t_0=0, \hat{\theta_0})$. A small error $\delta \theta^\prime$ in the measured lime temperature at $t^\prime$ results in a large change in the trajectory traced backwards, if $t^\prime$ is large enough to place the measured lime temperature close to the air temperature. See Fig.~\ref{fig:graphical_soln}.

\subsubsection{The prior distributions}
\paragraph{The initial temperature, $\Theta_0$.}
We impose a diffuse prior distribution on the initial temperature of the lime based on a (generous) range of temperatures encountered in refrigerators:
\begin{equation}
    \Theta_0\sim \mathcal{U}([0\,^\circ\text{C}, 15\,^\circ\text{C}]). \label{eq:theta_0_prior}
\end{equation}

\paragraph{The air temperature, $\Theta^{\text{air}}$.}
We impose an informative prior distribution on the air temperature $\Theta^{\text{air}}$ based on our (noisy) measurement of it:
\begin{equation}
    \Theta^{\text{air}} \sim \mathcal{N}(\theta_{\text{obs}}^{\text{air}}, \sigma^2).
\end{equation}

\paragraph{The parameter $\Lambda$ and variance of the measurement noise $\Sigma^2$.}
We exploit the information we obtained about $\Lambda$ and $\Sigma$ during our parameter identification activity in Sec.~\ref{sec:param_id} to construct informative prior distributions on the parameter $\Lambda$ characterizing the (same) lime and the variance in the measurement noise $\Sigma$ emanating from the (same) temperature sensor. Particularly, we use the posterior distributions from Sec.~\ref{sec:param_id}. ``Yesterday's posterior is today's prior.'' \cite{calvetti2010subjective}
\begin{subequations}
\begin{align}
    \Lambda & \sim \mathcal{N}(0.98\text{ hr}, (0.02\text{ hr})^2) \\
    \Sigma & \sim \mathcal{N}(0.16^\circ\text{C}, (0.03^\circ\text{C})^2).
\end{align}
\end{subequations}

\paragraph{The joint prior distribution.} Again, the joint prior distribution of all of the unknowns $\pi_{\text{prior}}(\theta_0, \lambda, \theta^{\text{air}}, \sigma)$ for this inverse problem factorizes since we impose independent priors.

\subsubsection{The data and likelihood function}
\paragraph{The data.} The data from the second heat transfer experiment are displayed in Fig.~\ref{fig:tr_data}:
\begin{itemize}
    \item the measured air temperature, $\theta_{\text{obs}}^{\text{air}}$
    \item a single measurement of the lime temperature, $(t^\prime, \theta_{\text{obs}}^\prime)$ with $t^\prime>t_0=0$.
\end{itemize} Note, the time the lime was taken out of the refrigerator $t_0=0$ is known.

\paragraph{The likelihood function.}
The likelihood function gives the probability density of the data $\theta_{\text{obs}}^\prime$ conditioned on each possible value of the parameters $\Lambda=\lambda$ and $\Sigma=\sigma$, and experimental conditions $\Theta_0=\theta_0$ and $\Theta^{\text{air}}=\theta^{\text{air}}$.
We construct the likelihood from the (i) data $(t^\prime, \theta_{\text{obs}}^\prime)$ and (ii) model of the data-generating process in eqn.~\ref{eq:Theta_obs}. 
The likelihood function is:
\begin{align}
   \pi_{\text{likelihood}}(\theta_{\text{obs}}^\prime \mid   \theta_0, \lambda,  \theta^{\text{air}}, \sigma) =
        \displaystyle 
        \frac{1}{\sigma \sqrt{2 \pi}}  
        \exp \left[-\frac{1}{2}\left(\frac{\theta_{ \text{obs}}^\prime-\theta(t^\prime ; \lambda, t_0=0, \theta_0, \theta^{\text{air}})}{\sigma}\right)^2 \right].  
    \label{eq:likelihood_tr}
\end{align}
It quantifies the support the datum $(t^\prime, \theta_{\text{obs}}^\prime)$ lends to each value of the unknowns $(\theta_0, \lambda,  \theta^{\text{air}}, \sigma)$.

\subsubsection{The posterior distribution}
The (joint) \emph{posterior density} governs the probability distribution of the unknowns $(\Theta_0, \Lambda, \Theta^{\text{air}}, \Sigma)$ conditioned on the data \thedatatr. 
By Bayes' theorem, the posterior density is proportional to the product of the likelihood function and prior density:
\begin{equation}
    \pi_{\text{posterior}}(\theta_0, \lambda, \theta^{\text{air}}, \sigma \mid \theta_{\text{obs}}^\prime) \propto 
    \pi_{\text{likelihood}}( \theta_{\text{obs}}^\prime \mid \theta_0, \lambda, \theta^{\text{air}}, \sigma)\pi_{\text{prior}}( \theta_0, \lambda, \theta^{\text{air}}, \sigma).
   \label{eq:posterior_tr}
\end{equation}

We are particularly interested in the posterior distribution of the initial lime temperature $\Theta_0$, with $(\Lambda, \Theta^{\text{air}}, \Sigma)$ marginalized out.

Again, we employ NUTS to obtain samples from the posterior then construct an empirical approximation of it. 

\subsubsection{Summary of results}
Fig.~\ref{fig:tr_data} shows the data from the heat transfer experiment that we employ to infer the initial lime temperature $\Theta_0$ with BSI.

\rocketemoji Fig.~\ref{fig:tr_prior_posterior} compares (i) the prior distribution of the initial lime temperature $\Theta_0$ with (ii) its updated, (marginal) empirical posterior distribution constructed via kernel density estimation. The bar shows the 90\% equal-tailed posterior credible interval for $\Theta_0$.  
Notably, the hold-out test data, the measured initial lime temperature $\theta_{0, \text{obs}}$, falls in this credible interval.

Fig.~\ref{fig:tr_trajectory} illustrates the posterior distribution of backward trajectories of the lime temperature by showing a random sample of 100 realizations of models for the lime temperature, $\theta(t;\lambda, t_0=0, \theta_0, \theta^{\text{air}})$, with $( \theta_0, \lambda, \theta^{\text{air}})$ a sample from the posterior distribution. 

\begin{figure}[h!]
    \centering
    \includegraphics[width=0.6\textwidth]{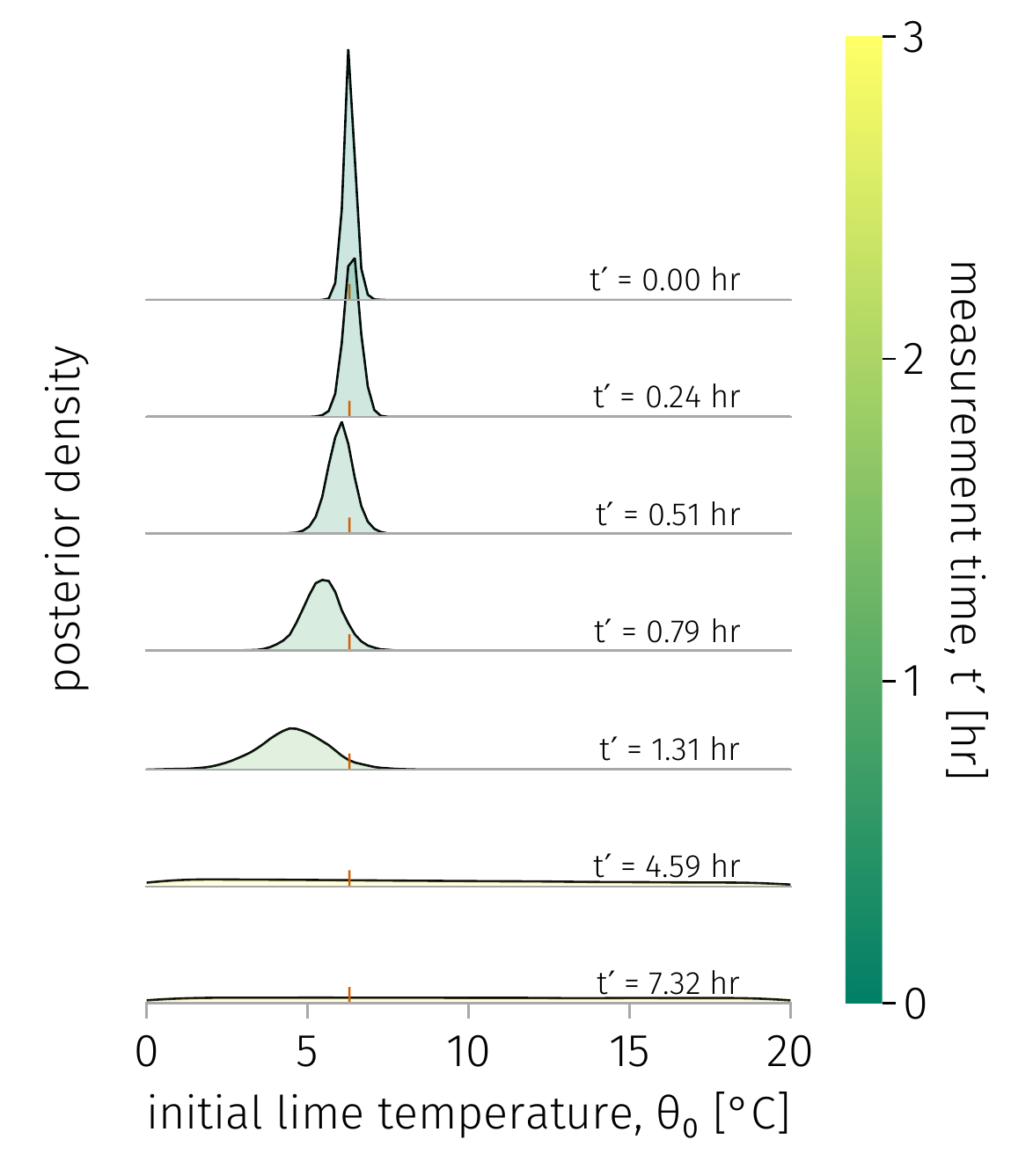} 
    \caption{\textbf{Inverse problem IIa: time reversal}: the (empirical, marginal) posterior distribution of $\Theta_0$ as the time at which we measure the lime temperature, $t^\prime$, increases.}
    \label{fig:tr_ridge}
\end{figure}

\paragraph{Ill-conditioning.} Fig.~\ref{fig:tr_ridge} shows the marginal posterior distribution of the initial lime temperature $\Theta_0$ for various times $t^\prime$ at which we measure the lime temperature to obtain \thedatatr. As $t^\prime$ becomes larger, the posterior distribution of $\Theta_0$ spreads, reflecting higher uncertainty. This illustrates the ability of BSI to capture ill-conditioning in inverse problems of reconstruction. 

\subsection{Inverse problem IIb: time reversal}
\label{sec:time_reversal_b}

\begin{mybox}[ 
floatplacement=t,
breakable,arc=0mm,
colback=cool_gray!10,
colframe=cool_green, 
fonttitle=\bfseries,
before upper={\parindent15pt}
]{Overview of problem and approach}
\small 
\noindent 
\textbf{Task}: employ BSI to infer both the time $T_0$ (ie., $t_0$ treated as a random variable) the lime was taken out of the refrigerator and the initial temperature $\Theta_0$ of the lime based on a measurement of the lime temperature at time $t^\prime>t_0$, $\theta_{\text{obs}}^\prime$, and the measured air temperature $\theta_{\text{obs}}^{\text{air}}$. \\

First, we impose a diffuse prior distribution on $\Theta_0$ based on the range of temperatures encountered in refrigerators and a weakly informative prior distribution on $T_0$ based on our sense of time passing. \\

Second, we impose informative prior distributions on the parameter $\Lambda$ characterizing the lime and the variance of the measurement noise $\Sigma^2$, based on the posterior distributions from our parameter identification activity in Sec.~\ref{sec:param_id}. \\

Next, we conduct another heat transfer experiment\footnote{Well, we use the same data from the second heat transfer experiment in Sec.~\ref{sec:time_reversal_a}. Importantly, it is a different experiment from the one used for parameter identification.} on the same lime (see Sec.~\ref{sec:setup}). Our measurement defining the condition of the experiment is the measured air temperature, $\theta^{\text{air}}_\text{obs}$. To (indirectly) gather information about $(T_0, \Theta_0)$, we measure the temperature of the lime at time $t^\prime$, $\theta^\prime_{\text{obs}}$. \\

Finally, we use Bayes' theorem to construct the posterior distribution of $(T_0, \Theta_0, \Lambda, \Theta^{\text{air}}, \Sigma)$ in light of the data, sample from it using a MCMC algorithm, and obtain an empirical marginal joint posterior distribution for the initial lime temperature $\Theta_0$ and time it was taken out of the refrigerator $T_0$ that quantifies posterior uncertainty about their values. \\

Note, we also measured the initial temperature of the lime and know when it was taken out of the refrigerator, but we hold this data $(t_0=0, \theta_{0, \text{obs}})$ out from the BSI procedure to test the fidelity of the posterior distribution of $(T_0, \Theta_0)$. \\ 

\textbf{Quick overview}:
\begin{itemize}
    \item \emph{data}: the measured air temperature $\theta^{\text{air}}_\text{obs}$ and lime temperature $(t^\prime, \theta_{\text{obs}}^\prime)$.
    \item \emph{random variables to infer from the data:} the time the lime was taken out of the refrigerator $T_0$, the initial lime temperature $\Theta_0$, the parameter $\Lambda$, the air temperature $\Theta^{\text{air}}$, the variance of the measurement noise $\Sigma^2$.
    \item \emph{sources of priors}: 
    $T_0$: our human judgement of the passing of time;
    $\Theta_0$: range of temperatures encountered in refrigerators;
    $\Lambda, \Sigma^2$: the posterior from our parameter identification activity in Sec.~\ref{sec:param_id}; 
    $\Theta^{\text{air}}$: our noisy measurement of the air temperature.
\end{itemize}
\end{mybox}

Classically, this time reversal problem is underdetermined. Conceptually, the current condition of the lime $(t^\prime, \theta_{\text{obs}}^\prime)$ is consistent with (1) ``the lime was initially very cold and has been outside of the refrigerator for a long duration'' \emph{and} (2) ``the lime was initially not very cold and has been outside of the refrigerator for a short duration''. 
Mathematically, there is a line of infinite solutions in the $(t_0, \theta_0)$ plane (the two primary unknowns) that satisfy the model in eqn.~\ref{eq:model} with known $(t^\prime, \theta^\prime)$ and $\theta^{\text{air}}$:
\begin{equation}
    \theta^\prime =\theta^{\text{air}}+(\theta_0-\theta^{\text{air}})e^{-(t^\prime-t_0)/\lambda} \label{eq:underdetermined_curve}.
\end{equation} 
Contrasting the classical curve of solutions in eqn.~\ref{eq:underdetermined_curve} with a solution via BSI, 
(i) depending on the prior distribution, the posterior distribution in BSI may assign different weights to each of the (inherently, equal-weighted) classical solutions and 
(ii) by accounting for measurement noise, BSI entertains solutions off of the curve comprising the classical solutions. 
This time reversal problem still becomes ill-conditioned for large $t^\prime$, in that the curve described by eqn.~\ref{eq:underdetermined_curve} becomes sensitive to errors in the measured $\theta^\prime$ as $t^\prime>t_0$ increases.

\begin{figure}[h!]
    \centering
    \centering
    \begin{subfigure}{0.48\textwidth}
        \includegraphics[width=\textwidth]{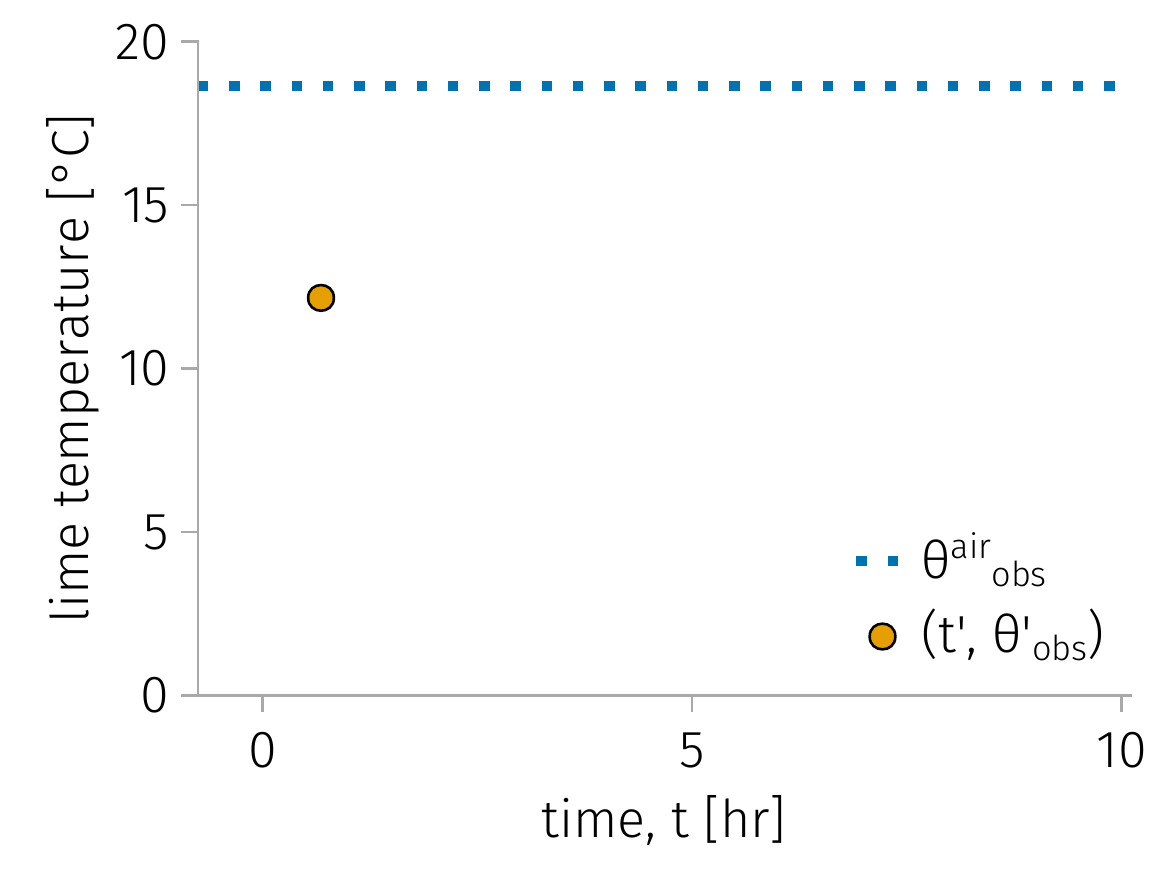} \caption{} \label{fig:tr2_data}
    \end{subfigure}

    \begin{subfigure}{0.48\textwidth}
        \includegraphics[width=\textwidth]{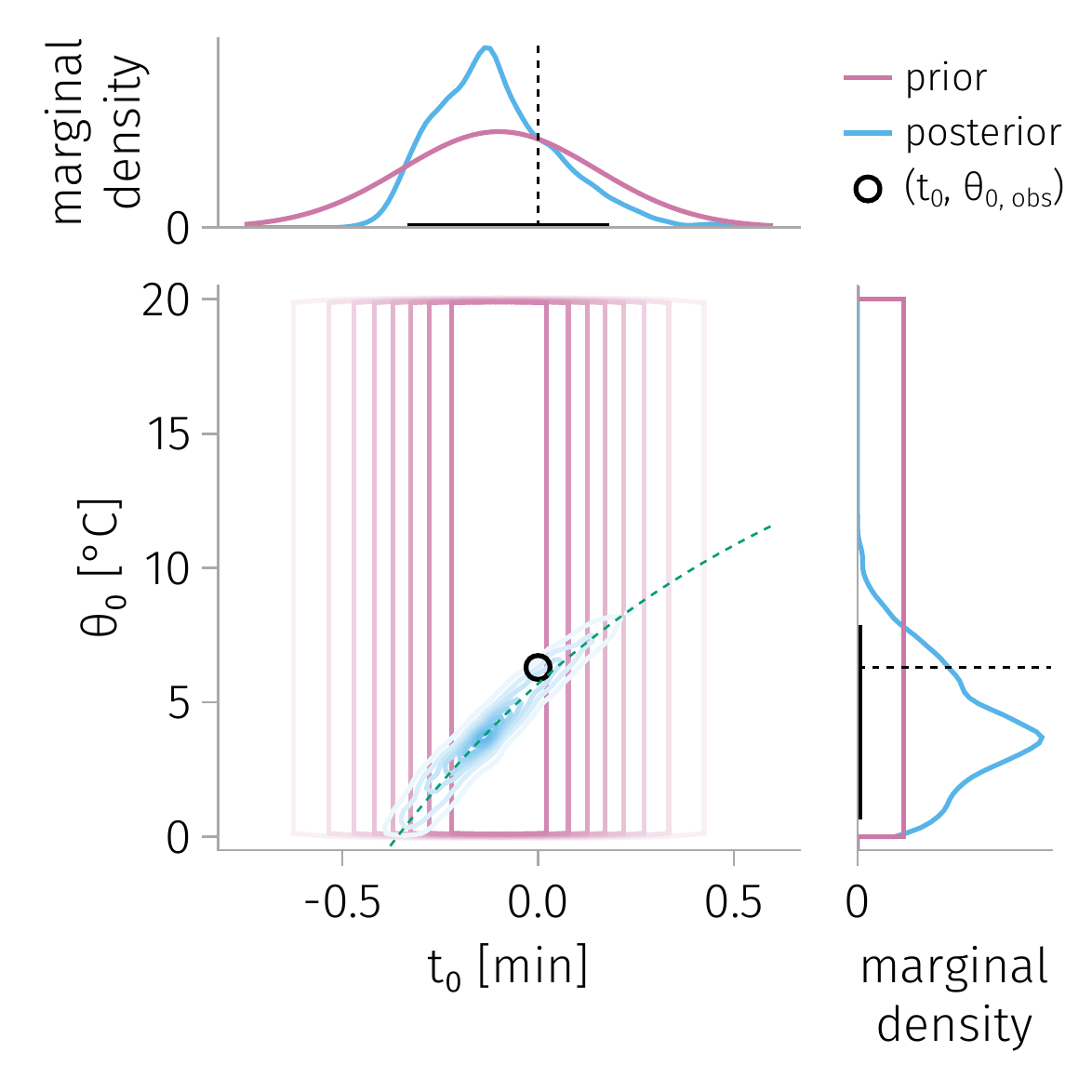} \caption{} \label{fig:tr2_prior_posterior}
    \end{subfigure}
    \begin{subfigure}{0.48\textwidth}
        \includegraphics[width=\textwidth]{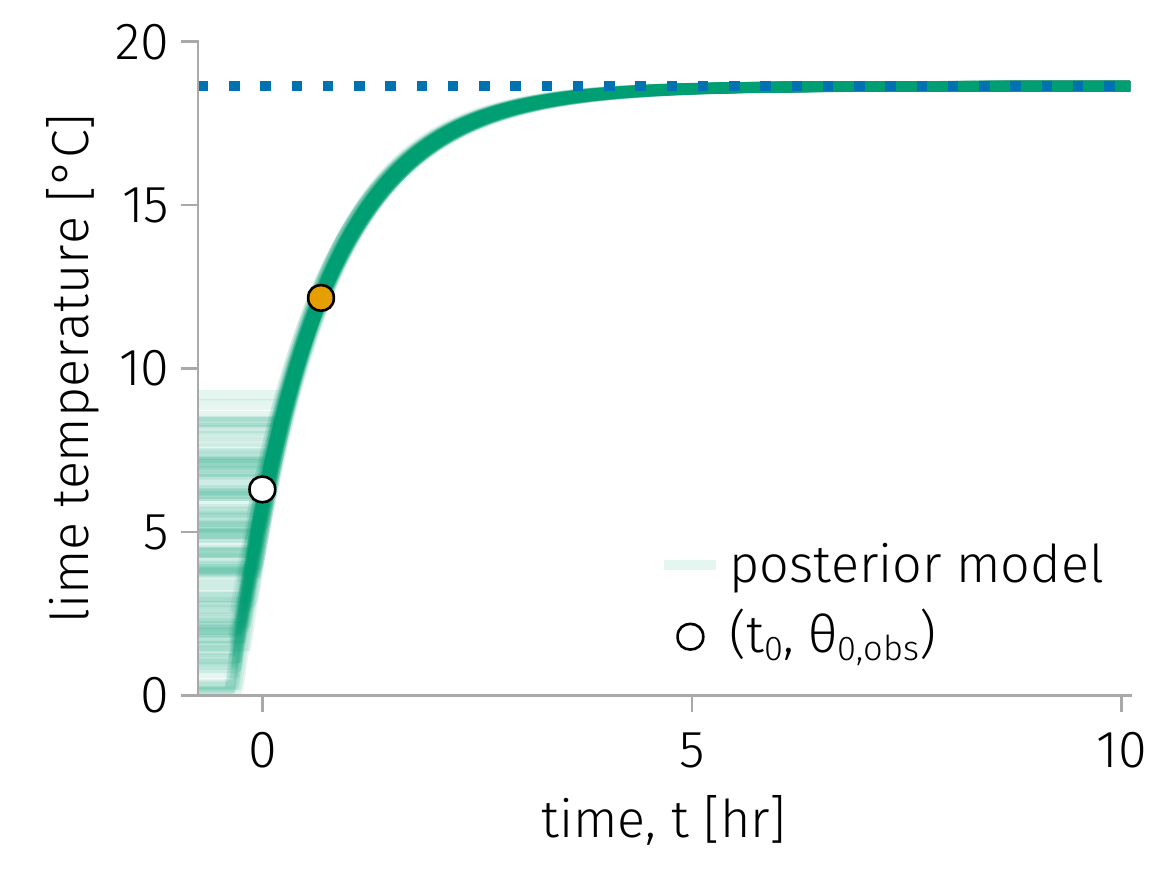} \caption{} \label{fig:tr2_trajectory}
    \end{subfigure}
    \caption{\textbf{Inverse problem IIb: time reversal}: infer the initial temperature $\Theta_0$ of the lime and the time $T_0$ it was taken out of the refrigerator from a measurement of its temperature later, $(t^\prime, \theta_{\text{obs}}^\prime)$.
    (a) The data.
    (b) The joint and marginal prior and posterior distributions of $(T_0, \Theta_0)$. The held-out measurement of the initial condition of the lime is indicated by the black point/dashed lines. The green dashed line shows the classical solution in eqn.~\ref{eq:underdetermined_curve} with $\lambda$ set to be the mean of the posterior from Sec.~\ref{sec:param_id}. The black bars in the marginal plots show the 90\% equal-tailed credible intervals.
    (c) A sample of 100 model trajectories $\theta(t;\lambda, t_0, \theta_0, \theta^{\text{air}})$, with $(T_0, \Theta_0, \Lambda, \Theta^{\text{air}})$ a sample from the posterior.
    }
\end{figure}

\subsubsection{The prior distributions}
We impose the same prior distributions of $\Theta_0$, $\Theta^{\text{air}}$, $\Lambda$, and $\Sigma^2$ as in Sec.~\ref{sec:time_reversal_a}. Additionally, we now impose a prior distribution on the time $T_0$ at which the lime was taken out of the refrigerator, based on our unreliable judgement of the passing of time: 
\begin{equation}
    T_0 \sim \mathcal{N}_{-1\leq T_0\leq 1}\left(-0.1\,\text{hr}, (0.25\,\text{hr})^2\right). \label{eq:T0_prior}
\end{equation}
This gives a joint prior distribution $\pi_{\text{prior}}(t_0, \theta_0, \lambda, \theta^{\text{air}}, \sigma)$ for this inverse problem. We visualize the prior of $(T_0, \Theta_0)$ in Fig.~\ref{fig:tr2_prior_posterior}.

\subsubsection{The data and likelihood function}
\paragraph{The data.} The data from the second heat transfer experiment are displayed in Fig.~\ref{fig:tr2_data}:
\begin{itemize}
    \item the measured air temperature, $\theta_{\text{obs}}^{\text{air}}$
    \item a single measurement of the lime temperature, $(t^\prime, \theta_{\text{obs}}^\prime)$ with $t^\prime$.
\end{itemize} The time the lime was taken out of the refrigerator $t_0$ is \emph{not} known.

\paragraph{The likelihood function.}
The likelihood function gives the probability density of the measurement $ \theta_{\text{obs}}^\prime$ conditioned on each possible value of the parameters $\Lambda=\lambda$ and $\Sigma=\sigma$ and experimental conditions $T_0=t_0$, $\Theta_0=\theta_0$, and $\Theta^{\text{air}}=\theta^{\text{air}}$.
\begin{align}
   \pi_{\text{likelihood}}(\theta_{\text{obs}}^\prime \mid t_0,  \theta_0, \lambda,  \theta^{\text{air}}, \sigma) =
        \displaystyle 
        \frac{1}{\sigma \sqrt{2 \pi}}  
        \exp \left[-\frac{1}{2}\left(\frac{\theta_{ \text{obs}}^\prime-\theta(t^\prime ; \lambda, t_0, \theta_0, \theta^{\text{air}})}{\sigma}\right)^2 \right].  
    \label{eq:likelihood_tr2}
\end{align}

\subsubsection{The posterior distribution}
The (joint) \emph{posterior density} governs the probability distribution of the unknowns $(T_0, \Theta_0, \Lambda, \Theta^{\text{air}}, \Sigma)$ conditioned on the data \thedatatr. 
By Bayes' theorem, the posterior density is proportional to the product of the likelihood function and (joint) prior density:
\begin{equation}
    \pi_{\text{posterior}}(t_0, \theta_0, \lambda, \theta^{\text{air}}, \sigma \mid \theta_{\text{obs}}^\prime) \propto 
    \pi_{\text{likelihood}}(\theta_{\text{obs}}^\prime \mid t_0, \theta_0, \lambda, \theta^{\text{air}}, \sigma)\pi_{\text{prior}}(t_0, \theta_0, \lambda, \theta^{\text{air}}, \sigma).
   \label{eq:posterior_tr2}
\end{equation}

We are particularly interested in the posterior distribution of the initial condition of the lime $(T_0, \Theta_0)$, with $(\Lambda, \Theta^{\text{air}}, \Sigma)$ marginalized out.

Again, we employ NUTS to obtain samples from the posterior. 

\subsubsection{Summary of results}
Fig.~\ref{fig:tr2_data} shows the data from the heat transfer experiment that we employ to infer the initial condition of the lime $(T_0, \Theta_0)$ with BSI.

\rocketemoji By showing contours, Fig.~\ref{fig:tr2_prior_posterior} compares (i) the joint prior distribution of the initial condition of the lime, $(T_0, \Theta_0)$, with (ii) the updated, empirical, joint posterior distribution of $(T_0, \Theta_0)$ constructed via kernel density estimation. Because this inverse problem is underdetermined, the posterior density is quite widely spread over the curve in the $(t_0, \theta_0)$ plane comprising the classical solution, in eqn.~\ref{eq:underdetermined_curve}. 
However, owing to the non-uniform density and finite support of the prior distribution of $(T_0, \Theta_0)$, generally each of the classical solutions is weighted differently in the posterior density. 
Also, the posterior density spreads orthogonal to the curve of classical solutions, quantifying uncertainty owing to admission of measurement noise in the data. 
Note, the hold-out test data, the measured initial condition of the lime $(t_0=0, \theta_{0, \text{obs}})$, falls in a region of high posterior density. But, this result is in part a consequence of the mean of our prior for $T_0$ in eqn.~\ref{eq:T0_prior} being close to the true $t_0=0$. The marginal posterior distributions of $T_0$ and $\Theta_0$ are compared in Fig.~\ref{fig:tr2_prior_posterior} as well, including their 90\%, equal-tailed posterior credible intervals.
The credible interval for $\Theta_0$ is much wider here than in the inverse problem IIa in Fig.~\ref{fig:tr_prior_posterior}, despite using the same measurement $\theta_{\text{obs}}^\prime$, because $t_0$ is not specified in this underdetermined inverse problem. 

Fig.~\ref{fig:tr2_trajectory} illustrates the posterior distribution of backward trajectories of the lime temperature by showing a random sample of 100 realizations of models for the lime temperature, $\theta(t;\lambda, t_0, \theta_0, \theta^{\text{air}})$, with $(t_0, \theta_0,\lambda, \theta^{\text{air}})$ a sample from the posterior distribution. The intuition explaining the correlation of $T_0$ and $\Theta_0$ the joint posterior density in Fig.~\ref{fig:tr2_prior_posterior} is apparent in Fig.~\ref{fig:tr2_trajectory}: the data $(t^\prime, \theta_{\text{obs}}^\prime)$ is consistent with both propositions (i) ``the lime was initially not very cold and taken out of the refrigerator recently'' and (ii) ``the lime was initially very cold and taken out of the refrigerator a while ago''.

\subsection{Conclusions}
By way of example, we provided a tutorial of Bayesian statistical inversion (BSI) to solve inverse problems while incorporating prior information and quantifying uncertainty. 
Our focus was a simple, intuitive physical process---heat transfer from ambient indoor air to a cold lime fruit via natural convection. 
First, we developed a simple mathematical model for the lime temperature, which contains a single parameter. 
Then, we used a time series data set of the lime temperature to infer, via BSI, the posterior distribution of the model parameter. 
Next, we employed the model with the inferred parameter to tackle, via BSI, two reconstruction problems of time reversal. The first task, ill-conditioned, was to predict the initial temperature of the lime from a measurement of its temperature later in time. The second task, underdetermined, was to predict the initial temperature of the lime \emph{and} the time it was taken out of the refrigerator from a measurement of its temperature later in time.
We intend for our tutorial to facilitate scientists and engineers to (i) recognize inverse problems in their domain and (ii) employ BSI to solve them, while incorporating prior information and quantifying uncertainty.

Our BSI solutions to the inverse problems involving the lime are subject to limitations. 
First, our mathematical model of the lime temperature relies on several simplifying assumptions listed in Sec.~\ref{sec:model}.
The model may be more accurate if we relaxed these assumptions and amended it to account for eg., 
(i) the time-dependence of the bulk air temperature $\theta^{\text{air}}(t)$ and
(ii) the spatial temperature gradients in the lime and its geometry and spatial heterogeneity (skin, flesh, seeds). 
Second, ideally, we would replicate the heat transfer experiment multiple times and use all of this time series data of the lime temperature for the inference of the parameter $\Lambda$ in Sec.~\ref{sec:param_id}. 
This would allow the posterior distribution of $\Sigma$ to capture the full residual variability of the lime temperature over repeated experiments, owing to poorly-controlled and/or unrecognized inputs/conditions that affect the lime temperature. 
Third, we neglected the possibility of model discrepancy \cite{kennedy2001bayesian} that could arise due to these factors, an unaccounted-for source of both uncertainty and systemic bias in the posterior of the parameter/initial condition of the lime.

\section{Discussion}
We provided an introductory tutorial on Bayesian statistical inversion as a tool to tackle inverse problems, which could be ill-posed, while (i) incorporating prior information and (ii) providing a solution in the form of a probability density function over the input/parameter space, which quantifies uncertainty via its spread. 
Inverse problems are pervasive throughout the sciences and engineering, eg.\ in heat or radiation transfer, gravitational fields, wave scattering, tomography, and electromagnetism \cite{kaipio2011bayesian,wang2004bayesian,orlande2014accelerated,isakov2006inverse,hasanouglu2021introduction,richter2021inverse}, vibration of springs and beams \cite{gladwell1986inverse}, imaging \cite{ribes2008linear,neto2012introduction}, fluid mechanics \cite{cotter2009bayesian}, physiology \cite{zenker2007inverse}, epidemiology \cite{hao2020reconstruction,ansari2022inferring}, ecology \cite{dowd2003bayesian}, geophysics \cite{richter2021inverse}, environmental science \cite{andrle2015inverse,yee2008bayesian}, palaeoclimatology \cite{haslett2006bayesian}, chemical/bio-chemical reactions \cite{engl2009inverse,guzzi2018inverse,santosa2011inverse,mebane2013bayesian,kugler2009parameter}, and adsorption \cite{shih2020hierarchical}.
Mathematically, reconstruction problems in these domains often reduce to using data to determine
(1) a vector that was linearly transformed by a matrix \cite{hofinger2007convergence,kaipio2006statistical,mohammad1996full},
(2) the initial condition, boundary condition, or forcing term in an ordinary or partial differential equation \cite{neto2012introduction}, 
(3) a function in an integrand \cite{groetsch2007integral,groetsch1993inverse},
or
(4) the geometry of a domain \cite{ito2014inverse}.

\paragraph{Classical approaches to uncertainty quantification.}
For overdetermined inverse problems tackled via least squares, bootstrapping \cite{efron1986bootstrap}, asymptotic theory \cite{banks2010standard}, and the Hessian of the loss function \cite{zhan2022uncertainty} can provide uncertainty estimates for the unknown input/parameters and new predictions by the model.

\paragraph{Model discrepancy.}
\emph{Model discrepancy} \cite{kennedy2001bayesian} refers to a nonzero difference between the (i) expected measurement of the system output over repeated experiments under the same conditions and (ii) the prediction of the system output by the model.
If significantly present and not accounted for in the model of the data-generating process, model discrepancy corrupts the BSI solution to an inverse problem \cite{brynjarsdottir2014learning}. 
First, disregarded model discrepancy introduces bias, ie.\ the posterior density of the parameter/input will not be centered at the true value even as more and more data are collected. 
Second, it leads to mis-calibrated uncertainty quantification, ie.\ the posterior credible interval will not be likely to contain the true value of the parameter/input even as more and more data are collected.

To account for model discrepancy, we can modify the model of the data-generating process to explicitly include a model discrepancy function $\Delta(x)$ (a random variable dependent on the input, $x$) as a Gaussian process \cite{seeger2004gaussian} and employ data to infer it. Ie., we model the measured system output $Y_{\text{obs}}$ as
\begin{equation}
    Y_{\text{obs}}(x) = f_\beta(x) + \Delta(x) + \Epsilon, \label{eq:Y_obs_model_disc}
\end{equation} with $\Epsilon$ the noise, which by definition has mean zero.

\paragraph{Model selection.} We assumed the model structure is known (see Tab.~\ref{tab:unify_framing}). 
In \emph{model selection}, the task is to select the best model of a physical system among a set of candidate models, given data. Bayesian approaches to model selection include the Bayesian information criterion \cite{neath2012bayesian} and Bayes factors \cite{kass1995bayes,van2021bayesian}. 


\paragraph{Other methods to approximate the posterior distribution.} 
Here, we used the NUTS \cite{hoffman2014no} Markov Chain Monte Carlo method to obtain samples from the posterior distribution to approximate it. 
Other methods to approximate the posterior distribution include 
(i) other Monte Carlo sampling methods, including Approximate Bayesian Computation (ABC) \cite{sunnaaker2013approximate}, sequential Monte Carlo \cite{murphy2023probabilistic}, and the adaptive Metropolis algorithm \cite{haario2001adaptive} and 
(ii) obtaining a tractable analytical expression as an approximation to the posterior, including the Laplace approximation \cite{sun2013review} and variational Bayes \cite{sun2013review}.
The BSI practitioner must consider accuracy, speed, and ease of implementation \cite{murphy2023probabilistic} when choosing a method to approximate their posterior at hand.


\section*{Data and code availability}
All data and Julia code to reproduce the plots in this article are available at \url{github.com/faaiqgwaqar/Inverse-Problems}. 

\section*{Acknowledgements}
F.W. acknowledges NSF award 1920945 for support.
C.M.S. acknowledges support from the US Department of Homeland Security Countering Weapons of Mass Destruction under CWMD Academic Research Initiative Cooperative Agreement 21CWDARI00043. This support does not constitute an expressed or implied endorsement on the part of the Government.
Thanks to Edward Celarier and Luther Mahoney for feedback on our manuscript.

\clearpage

\bibliography{refs}

\begin{thebibliography}{10}

\bibitem{epstein2008model}
Joshua~M Epstein.
\newblock Why model?
\newblock {\em Journal of Artificial Societies and Social Simulation},
  11(4):12, 2008.

\bibitem{aster2018parameter}
Richard~C Aster, Brian Borchers, and Clifford~H Thurber.
\newblock {\em Parameter estimation and inverse problems}.
\newblock Elsevier, 2018.

\bibitem{groetsch1999inverse}
Charles~W Groetsch.
\newblock {\em Inverse problems: activities for undergraduates}, volume~12.
\newblock Cambridge University Press, 1999.

\bibitem{iglesias2014inverse}
Marco Iglesias and Andrew~M Stuart.
\newblock Inverse problems and uncertainty quantification.
\newblock {\em SIAM News}, 20:2--3, 2014.

\bibitem{strogatz2018nonlinear}
Steven~H Strogatz.
\newblock {\em Nonlinear dynamics and chaos: with applications to physics,
  biology, chemistry, and engineering}.
\newblock CRC press, 2018.

\bibitem{kabanikhin2008definitions}
Sergei~Igorevich Kabanikhin.
\newblock Definitions and examples of inverse and ill-posed problems.
\newblock {\em Journal of Inverse and Ill-Posed Problems}, 16(4):317--357,
  2008.

\bibitem{neto2012introduction}
Francisco Duarte~Moura Neto and Ant{\^o}nio~Jos{\'e} da~Silva~Neto.
\newblock {\em An introduction to inverse problems with applications}.
\newblock Springer Science \& Business Media, 2012.

\bibitem{fernandez2013bayes}
Juan~Luis Fern{\'a}ndez-Mart{\'\i}nez, Z~Fern{\'a}ndez-Mu{\~n}iz, JLG Pallero,
  and Luis~Mariano Pedruelo-Gonz{\'a}lez.
\newblock {From Bayes to Tarantola: new insights to understand uncertainty in
  inverse problems}.
\newblock {\em Journal of Applied Geophysics}, 98:62--72, 2013.

\bibitem{ito2014inverse}
Kazufumi Ito and Bangti Jin.
\newblock {\em Inverse problems: Tikhonov theory and algorithms}, volume~22.
\newblock World Scientific, 2014.

\bibitem{denisov1999elements}
Aleksandr~Mihajlovi{\v{c}} Denisov.
\newblock {\em Elements of the theory of inverse problems}, volume~14.
\newblock VSP, 1999.

\bibitem{mueller2012linear}
Jennifer~L Mueller and Samuli Siltanen.
\newblock {\em Linear and nonlinear inverse problems with practical
  applications}.
\newblock SIAM, 2012.

\bibitem{tenorio2017introduction}
Luis Tenorio.
\newblock {\em An introduction to data analysis and uncertainty quantification
  for inverse problems}.
\newblock SIAM, 2017.

\bibitem{kaipio2006statistical}
Jari Kaipio and Erkki Somersalo.
\newblock {\em Statistical and computational inverse problems}, volume 160.
\newblock Springer Science \& Business Media, 2006.

\bibitem{heilio2016mathematical}
Marko Vauhkonen, Tanja Tarvainen, and Timo L\"ahivaara.
\newblock Inverse problems.
\newblock In {\em Mathematical modelling}. Springer, 2016.

\bibitem{kac1966can}
Mark Kac.
\newblock Can one hear the shape of a drum?
\newblock {\em The American Mathematical Monthly}, 73(4P2):1--23, 1966.

\bibitem{protter1987can}
MH~Protter.
\newblock {Can one hear the shape of a drum? Revisited}.
\newblock {\em SIAM Review}, 29(2):185--197, 1987.

\bibitem{gordon1992one}
Carolyn Gordon, David~L Webb, and Scott Wolpert.
\newblock One cannot hear the shape of a drum.
\newblock {\em Bulletin of the American Mathematical Society}, 27(1):134--138,
  1992.

\bibitem{gordon1996you}
Carolyn Gordon and David Webb.
\newblock You can't hear the shape of a drum.
\newblock {\em American Scientist}, 84(1):46--55, 1996.

\bibitem{kreyszig2009advanced}
Erwin Kreyszig.
\newblock Advanced engineering mathematics, 10th edition, 2009.

\bibitem{kuttler1984eigenvalues}
James~R Kuttler and Vincent~G Sigillito.
\newblock {Eigenvalues of the Laplacian in two dimensions}.
\newblock {\em SIAM Review}, 26(2):163--193, 1984.

\bibitem{chavent2010nonlinear}
Guy Chavent.
\newblock {\em Nonlinear least squares for inverse problems: theoretical
  foundations and step-by-step guide for applications}.
\newblock Springer Science \& Business Media, 2010.

\bibitem{lines1984review}
LR~Lines and S~Treitel.
\newblock A review of least-squares inversion and its application to
  geophysical problems.
\newblock {\em Geophysical Prospecting}, 32(2):159--186, 1984.

\bibitem{kennedy2001bayesian}
Marc~C Kennedy and Anthony O'Hagan.
\newblock Bayesian calibration of computer models.
\newblock {\em Journal of the Royal Statistical Society: Series B (Statistical
  Methodology)}, 63(3):425--464, 2001.

\bibitem{sabatier2000past}
Pierre~C Sabatier.
\newblock Past and future of inverse problems.
\newblock {\em Journal of Mathematical Physics}, 41(6):4082--4124, 2000.

\bibitem{maclaren2019can}
Oliver~J Maclaren and Ruanui Nicholson.
\newblock {What can be estimated? Identifiability, estimability, causal
  inference and ill-posed inverse problems}.
\newblock {\em arXiv preprint arXiv:1904.02826}, 2019.

\bibitem{guillaume2019introductory}
Joseph~HA Guillaume, John~D Jakeman, Stefano Marsili-Libelli, Michael Asher,
  Philip Brunner, Barry Croke, Mary~C Hill, Anthony~J Jakeman, Karel~J Keesman,
  Saman Razavi, and Johannes~D Stigterers.
\newblock Introductory overview of identifiability analysis: A guide to
  evaluating whether you have the right type of data for your modeling purpose.
\newblock {\em Environmental Modelling \& Software}, 119:418--432, 2019.

\bibitem{chis2011structural}
Oana-Teodora Chis, Julio~R Banga, and Eva Balsa-Canto.
\newblock Structural identifiability of systems biology models: a critical
  comparison of methods.
\newblock {\em PloS One}, 6(11):e27755, 2011.

\bibitem{dashti2017bayesian}
Masoumeh Dashti and Andrew~M Stuart.
\newblock {The Bayesian approach to inverse problems}.
\newblock In {\em Handbook of uncertainty quantification}, pages 311--428.
  Springer, 2017.

\bibitem{tarantola1982inverse}
Albert Tarantola, Bernard Valette, et~al.
\newblock Inverse problems= quest for information.
\newblock {\em Journal of geophysics}, 50(1):159--170, 1982.

\bibitem{stuart2010inverse}
Andrew~M Stuart.
\newblock Inverse problems: a bayesian perspective.
\newblock {\em Acta Numerica}, 19:451--559, 2010.

\bibitem{tarantola2005inverse}
Albert Tarantola.
\newblock {\em Inverse problem theory and methods for model parameter
  estimation}.
\newblock SIAM, 2005.

\bibitem{idier2013bayesian}
J{\'e}r{\^o}me Idier.
\newblock {\em Bayesian approach to inverse problems}.
\newblock John Wiley \& Sons, 2013.

\bibitem{ulrych2001bayes}
Tadeusz~J Ulrych, Mauricio~D Sacchi, and Alan Woodbury.
\newblock {A Bayes tour of inversion: A tutorial}.
\newblock {\em Geophysics}, 66(1):55--69, 2001.

\bibitem{fitzpatrick1991bayesian}
Ben~G Fitzpatrick.
\newblock Bayesian analysis in inverse problems.
\newblock {\em Inverse Problems}, 7(5):675, 1991.

\bibitem{calvetti2018inverse}
Daniela Calvetti and Erkki Somersalo.
\newblock Inverse problems: From regularization to bayesian inference.
\newblock {\em Wiley Interdisciplinary Reviews: Computational Statistics},
  10(3):e1427, 2018.

\bibitem{trotta2008bayes}
Roberto Trotta.
\newblock Bayes in the sky: Bayesian inference and model selection in
  cosmology.
\newblock {\em Contemporary Physics}, 49(2):71--104, 2008.

\bibitem{ghosh2006introduction}
Jayanta~K Ghosh, Mohan Delampady, and Tapas Samanta.
\newblock {\em An introduction to Bayesian analysis: theory and methods},
  volume 725.
\newblock Springer, 2006.

\bibitem{van2021bayesian}
Rens van~de Schoot, Sarah Depaoli, Ruth King, Bianca Kramer, Kaspar
  M{\"a}rtens, Mahlet~G Tadesse, Marina Vannucci, Andrew Gelman, Duco Veen,
  Joukje Willemsen, and Christopher Yau.
\newblock Bayesian statistics and modelling.
\newblock {\em Nature Reviews Methods Primers}, 1(1):1--26, 2021.

\bibitem{downey2021think}
Allen~B Downey.
\newblock {Think Bayes 2}.
\newblock \url{https://allendowney.github.io/ThinkBayes2/index.html}, 2021.

\bibitem{murphy2023probabilistic}
Kevin~P Murphy.
\newblock {\em Probabilistic machine learning: Advanced topics}.
\newblock MIT Press, 2023.

\bibitem{hyndman1996computing}
Rob~J Hyndman.
\newblock Computing and graphing highest density regions.
\newblock {\em The American Statistician}, 50(2):120--126, 1996.

\bibitem{koch2007introduction}
Karl-Rudolf Koch.
\newblock {\em Introduction to Bayesian statistics}.
\newblock Springer Science \& Business Media, 2007.

\bibitem{hespanhol2019understanding}
Luiz Hespanhol, Caio~Sain Vallio, Luc{\'\i}ola~Menezes Costa, and Bruno~T
  Saragiotto.
\newblock Understanding and interpreting confidence and credible intervals
  around effect estimates.
\newblock {\em {Brazilian Journal of Physical Therapy}}, 23(4):290--301, 2019.

\bibitem{cengel2012ebook}
Yunus Cengel, John Cimbala, and Robert Turner.
\newblock {\em Fundamentals of Thermal-Fluid Sciences}.
\newblock McGraw Hill, 2017.

\bibitem{bird2002transport}
R~Byron Bird, Warren~E Stewart, and Edwin~N. Lightfoot.
\newblock {\em Transport phenomena}.
\newblock John Wiley and Sons, 2002.

\bibitem{ikegwu2009thermal}
OJ~Ikegwu, FC~Ekwu, et~al.
\newblock {Thermal and physical properties of some tropical fruits and their
  juices in Nigeria}.
\newblock {\em Journal of Food Technology}, 7(2):38--42, 2009.

\bibitem{KOSKY2013259}
Philip Kosky, Robert Balmer, William Keat, and George Wise.
\newblock Chapter 12 - mechanical engineering.
\newblock In Philip Kosky, Robert Balmer, William Keat, and George Wise,
  editors, {\em Exploring Engineering}, pages 259--281. Academic Press, Boston,
  3 edition, 2013.

\bibitem{vollmer2009newton}
Michael Vollmer.
\newblock Newton's law of cooling revisited.
\newblock {\em European Journal of Physics}, 30(5):1063, 2009.

\bibitem{rees1988cooling}
WG~Rees and C~Viney.
\newblock On cooling tea and coffee.
\newblock {\em American Journal of Physics}, 56(5):434--437, 1988.

\bibitem{o1990newton}
Colm~T O’Sullivan.
\newblock Newton’s law of cooling---a critical assessment.
\newblock {\em American Journal of Physics}, 58(10):956--960, 1990.

\bibitem{bohren1991comment}
Craig~F Bohren.
\newblock {Comment on ‘‘Newton’s law of cooling—A critical
  assessment,’’ by Colm T. O’Sullivan [Am. J. Phys. 5 8, 956--960
  (1990)]}.
\newblock {\em American Journal of Physics}, 59(11):1044--1046, 1991.

\bibitem{mukama2020thermophysical}
Matia Mukama, Alemayehu Ambaw, and Umezuruike~Linus Opara.
\newblock Thermophysical properties of fruit—a review with reference to
  postharvest handling.
\newblock {\em Journal of Food Measurement and Characterization},
  14(5):2917--2937, 2020.

\bibitem{hoffman2014no}
Matthew~D Hoffman, Andrew Gelman, et~al.
\newblock {The No-U-Turn sampler: adaptively setting path lengths in
  Hamiltonian Monte Carlo}.
\newblock {\em J. Mach. Learn. Res.}, 15(1):1593--1623, 2014.

\bibitem{ge2018t}
Hong Ge, Kai Xu, and Zoubin Ghahramani.
\newblock Turing: a language for flexible probabilistic inference.
\newblock In {\em International Conference on Artificial Intelligence and
  Statistics, {AISTATS} 2018, 9-11 April 2018, Playa Blanca, Lanzarote, Canary
  Islands, Spain}, pages 1682--1690, 2018.

\bibitem{bezanson2012julia}
Jeff Bezanson, Stefan Karpinski, Viral~B Shah, and Alan Edelman.
\newblock Julia: A fast dynamic language for technical computing.
\newblock {\em arXiv preprint arXiv:1209.5145}, 2012.

\bibitem{chen2017tutorial}
Yen-Chi Chen.
\newblock A tutorial on kernel density estimation and recent advances.
\newblock {\em Biostatistics \& Epidemiology}, 1(1):161--187, 2017.

\bibitem{sherlock2010random}
Chris Sherlock, Paul Fearnhead, and Gareth~O Roberts.
\newblock {The random walk Metropolis: linking theory and practice through a
  case study}.
\newblock {\em Statistical Science}, 25(2):172--190, 2010.

\bibitem{roy2020convergence}
Vivekananda Roy.
\newblock {Convergence diagnostics for Markov chain Monte Carlo}.
\newblock {\em Annual Review of Statistics and Its Application}, 7:387--412,
  2020.

\bibitem{roberts2001optimal}
Gareth~O Roberts and Jeffrey~S Rosenthal.
\newblock {Optimal scaling for various Metropolis-Hastings algorithms}.
\newblock {\em Statistical Science}, 16(4):351--367, 2001.

\bibitem{betancourt2017conceptual}
Michael Betancourt.
\newblock {A conceptual introduction to Hamiltonian Monte Carlo}.
\newblock {\em arXiv preprint arXiv:1701.02434}, 2017.

\bibitem{calvetti2010subjective}
D~Calvetti and E~Somersalo.
\newblock {Subjective knowledge or objective belief? An oblique look to
  Bayesian methods}.
\newblock {\em Large-Scale Inverse Problems and Quantification of Uncertainty},
  pages 33--70, 2010.

\bibitem{kaipio2011bayesian}
Jari~P Kaipio and Colin Fox.
\newblock {The Bayesian framework for inverse problems in heat transfer}.
\newblock {\em Heat Transfer Engineering}, 32(9):718--753, 2011.

\bibitem{wang2004bayesian}
Jingbo Wang and Nicholas Zabaras.
\newblock A bayesian inference approach to the inverse heat conduction problem.
\newblock {\em International Journal of Heat and Mass Transfer},
  47(17-18):3927--3941, 2004.

\bibitem{orlande2014accelerated}
Helcio~RB Orlande, George~S Dulikravich, Markus Neumayer, Daniel Watzenig, and
  Marcelo~J Cola{\c{c}}o.
\newblock {Accelerated Bayesian inference for the estimation of spatially
  varying heat flux in a heat conduction problem}.
\newblock {\em Numerical Heat Transfer, Part A: Applications}, 65(1):1--25,
  2014.

\bibitem{isakov2006inverse}
Victor Isakov.
\newblock {\em Inverse problems for partial differential equations}, volume
  127.
\newblock Springer, 2006.

\bibitem{hasanouglu2021introduction}
Alemdar~Hasanov Hasano{\u{g}}lu and Vladimir~G Romanov.
\newblock {\em Introduction to inverse problems for differential equations}.
\newblock Springer, 2021.

\bibitem{richter2021inverse}
Mathias Richter.
\newblock {\em Inverse problems: Basics, theory and applications in
  geophysics}.
\newblock Springer Nature, 2021.

\bibitem{gladwell1986inverse}
Graham~ML Gladwell.
\newblock {\em Inverse problems in vibration}.
\newblock 1986.

\bibitem{ribes2008linear}
Alejandro Ribes and Francis Schmitt.
\newblock Linear inverse problems in imaging.
\newblock {\em IEEE Signal Processing Magazine}, 25(4):84--99, 2008.

\bibitem{cotter2009bayesian}
Simon~L Cotter, Massoumeh Dashti, James~Cooper Robinson, and Andrew~M Stuart.
\newblock Bayesian inverse problems for functions and applications to fluid
  mechanics.
\newblock {\em Inverse Problems}, 25(11):115008, 2009.

\bibitem{zenker2007inverse}
Sven Zenker, Jonathan Rubin, and Gilles Clermont.
\newblock From inverse problems in mathematical physiology to quantitative
  differential diagnoses.
\newblock {\em PLoS Computational Biology}, 3(11):e204, 2007.

\bibitem{hao2020reconstruction}
Xingjie Hao, Shanshan Cheng, Degang Wu, Tangchun Wu, Xihong Lin, and Chaolong
  Wang.
\newblock {Reconstruction of the full transmission dynamics of COVID-19 in
  Wuhan}.
\newblock {\em Nature}, 584(7821):420--424, 2020.

\bibitem{ansari2022inferring}
Mehrad Ansari, David Soriano-Pa{\~n}os, Gourab Ghoshal, and Andrew~D White.
\newblock Inferring spatial source of disease outbreaks using maximum entropy.
\newblock {\em Physical Review E}, 106(1):014306, 2022.

\bibitem{dowd2003bayesian}
Michael Dowd and Renate Meyer.
\newblock {A Bayesian approach to the ecosystem inverse problem}.
\newblock {\em Ecological modelling}, 168(1-2):39--55, 2003.

\bibitem{andrle2015inverse}
M~Andrle and A~El~Badia.
\newblock {On an inverse source problem for the heat equation. Application to a
  pollution detection problem, II}.
\newblock {\em Inverse Problems in Science and Engineering}, 23(3):389--412,
  2015.

\bibitem{yee2008bayesian}
Eugene Yee, Fue-Sang Lien, Andrew Keats, and R{\'e}al D’Amours.
\newblock Bayesian inversion of concentration data: Source reconstruction in
  the adjoint representation of atmospheric diffusion.
\newblock {\em Journal of Wind Engineering and Industrial Aerodynamics},
  96(10-11):1805--1816, 2008.

\bibitem{haslett2006bayesian}
John Haslett, Matt Whiley, Sudipto Bhattacharya, M~Salter-Townshend, Simon~P
  Wilson, JRM Allen, B~Huntley, and FJG Mitchell.
\newblock Bayesian palaeoclimate reconstruction.
\newblock {\em Journal of the Royal Statistical Society: Series A (Statistics
  in Society)}, 169(3):395--438, 2006.

\bibitem{engl2009inverse}
Heinz~W Engl, Christoph Flamm, Philipp K{\"u}gler, James Lu, Stefan M{\"u}ller,
  and Peter Schuster.
\newblock Inverse problems in systems biology.
\newblock {\em Inverse Problems}, 25(12):123014, 2009.

\bibitem{guzzi2018inverse}
Rodolfo Guzzi, Teresa Colombo, and Paola Paci.
\newblock Inverse problems in systems biology: A critical review.
\newblock {\em Systems Biology}, pages 69--94, 2018.

\bibitem{santosa2011inverse}
Fadil Santosa and Benjamin Weitz.
\newblock An inverse problem in reaction kinetics.
\newblock {\em Journal of Mathematical Chemistry}, 49(8), 2011.

\bibitem{mebane2013bayesian}
David~S Mebane, K~Sham Bhat, Joel~D Kress, Daniel~J Fauth, McMahan~L Gray,
  Andrew Lee, and David~C Miller.
\newblock {Bayesian calibration of thermodynamic models for the uptake of
  CO$_2$ in supported amine sorbents using ab initio priors}.
\newblock {\em Physical Chemistry Chemical Physics}, 15(12):4355--4366, 2013.

\bibitem{kugler2009parameter}
Philipp Kugler, Erwin Gaubitzer, and Stefan M{\"u}ller.
\newblock Parameter identification for chemical reaction systems using sparsity
  enforcing regularization: A case study for the chlorite- iodide reaction.
\newblock {\em The Journal of Physical Chemistry A}, 113(12):2775--2785, 2009.

\bibitem{shih2020hierarchical}
Chunkai Shih, Jongwoo Park, David~S Sholl, Matthew~J Realff, Tomoyuki Yajima,
  and Yoshiaki Kawajiri.
\newblock Hierarchical bayesian estimation for adsorption isotherm parameter
  determination.
\newblock {\em Chemical Engineering Science}, 214:115435, 2020.

\bibitem{hofinger2007convergence}
Andreas Hofinger and Hanna~K Pikkarainen.
\newblock {Convergence rate for the Bayesian approach to linear inverse
  problems}.
\newblock {\em Inverse Problems}, 23(6):2469, 2007.

\bibitem{mohammad1996full}
Ali Mohammad-Djafari.
\newblock {A full Bayesian approach for inverse problems}.
\newblock In {\em Maximum Entropy and Bayesian Methods: Santa Fe, New Mexico,
  USA, 1995 Proceedings of the Fifteenth International Workshop on Maximum
  Entropy and Bayesian Methods}, pages 135--144. Springer, 1996.

\bibitem{groetsch2007integral}
Charles~W Groetsch.
\newblock Integral equations of the first kind, inverse problems and
  regularization: a crash course.
\newblock {\em Journal of Physics: Conference Series}, 73(1):012001, 2007.

\bibitem{groetsch1993inverse}
Charles~W Groetsch and CW~Groetsch.
\newblock {\em Inverse problems in the mathematical sciences}, volume~52.
\newblock Springer, 1993.

\bibitem{efron1986bootstrap}
Bradley Efron and Robert Tibshirani.
\newblock Bootstrap methods for standard errors, confidence intervals, and
  other measures of statistical accuracy.
\newblock {\em Statistical Science}, pages 54--75, 1986.

\bibitem{banks2010standard}
Harvey~Thomas Banks, Kathleen Holm, and Danielle Robbins.
\newblock Standard error computations for uncertainty quantification in inverse
  problems: Asymptotic theory vs. bootstrapping.
\newblock {\em Mathematical and Computer Modelling}, 52(9-10):1610--1625, 2010.

\bibitem{zhan2022uncertainty}
Ni~Zhan and John~R Kitchin.
\newblock Uncertainty quantification in machine learning and nonlinear least
  squares regression models.
\newblock {\em AIChE Journal}, 68(6):e17516, 2022.

\bibitem{brynjarsdottir2014learning}
Jenn{\`y} Brynjarsd{\'o}ttir and Anthony O'Hagan.
\newblock Learning about physical parameters: The importance of model
  discrepancy.
\newblock {\em Inverse Problems}, 30(11):114007, 2014.

\bibitem{seeger2004gaussian}
Matthias Seeger.
\newblock Gaussian processes for machine learning.
\newblock {\em International Journal of Neural Systems}, 14(02):69--106, 2004.

\bibitem{neath2012bayesian}
Andrew~A Neath and Joseph~E Cavanaugh.
\newblock {The Bayesian information criterion: background, derivation, and
  applications}.
\newblock {\em Wiley Interdisciplinary Reviews: Computational Statistics},
  4(2):199--203, 2012.

\bibitem{kass1995bayes}
Robert~E Kass and Adrian~E Raftery.
\newblock Bayes factors.
\newblock {\em Journal of the American Statistical Association},
  90(430):773--795, 1995.

\bibitem{sunnaaker2013approximate}
Mikael Sunn{\aa}ker, Alberto~Giovanni Busetto, Elina Numminen, Jukka Corander,
  Matthieu Foll, and Christophe Dessimoz.
\newblock {Approximate Bayesian computation}.
\newblock {\em PLoS Computational Biology}, 9(1):e1002803, 2013.

\bibitem{haario2001adaptive}
Heikki Haario, Eero Saksman, and Johanna Tamminen.
\newblock {An adaptive Metropolis algorithm}.
\newblock {\em Bernoulli}, pages 223--242, 2001.

\bibitem{sun2013review}
Shiliang Sun.
\newblock A review of deterministic approximate inference techniques for
  bayesian machine learning.
\newblock {\em Neural Computing and Applications}, 23:2039--2050, 2013.

\bibitem{wu2018basic}
Joseph Wu.
\newblock {A basic guide to RTD measurements}.
\newblock {\em Change}, 100(4.183):10--12, 2018.

\bibitem{king2004rtd}
Grayson King and Toru Fukushima.
\newblock {RTD Interfacing and Linearization Using an ADuC8xx
  MicroConverter{\textregistered}}.
\newblock {\em AN709, Analog Devices}, 2004.

\end{thebibliography}
\bibliographystyle{unsrt}

\clearpage

\renewcommand{\thepage}{S\arabic{page}}  
\renewcommand{\thesection}{S\arabic{section}}   
\renewcommand{\thetable}{S\arabic{table}}   
\renewcommand{\thefigure}{S\arabic{figure}}
\renewcommand{\theequation}{S\arabic{equation}}

\setcounter{section}{0}
\setcounter{figure}{0}   
\setcounter{page}{1}

\section{Supporting Information}

\subsection{The temperature sensor and Arduino setup}
\label{sec:temp_sensing}
\begin{figure}[h!]
\centering 
  \begin{subfigure}[b]{0.3\textwidth}
    \includegraphics[width=\columnwidth]{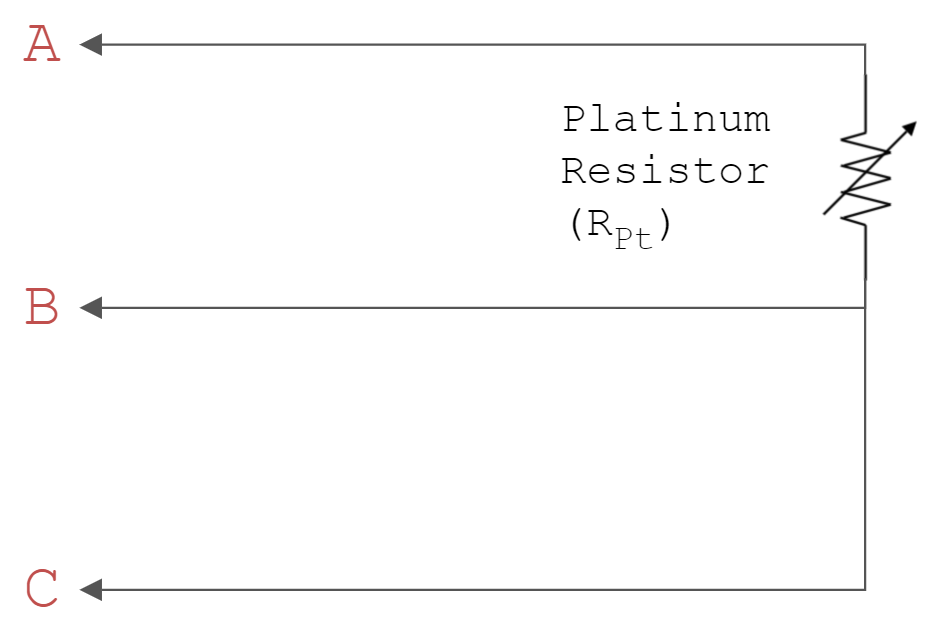} \caption{The RTD temperature sensor} \label{fig:rtd}
  \end{subfigure}
  \begin{subfigure}[b]{0.5\textwidth}
    \includegraphics[width=\columnwidth]{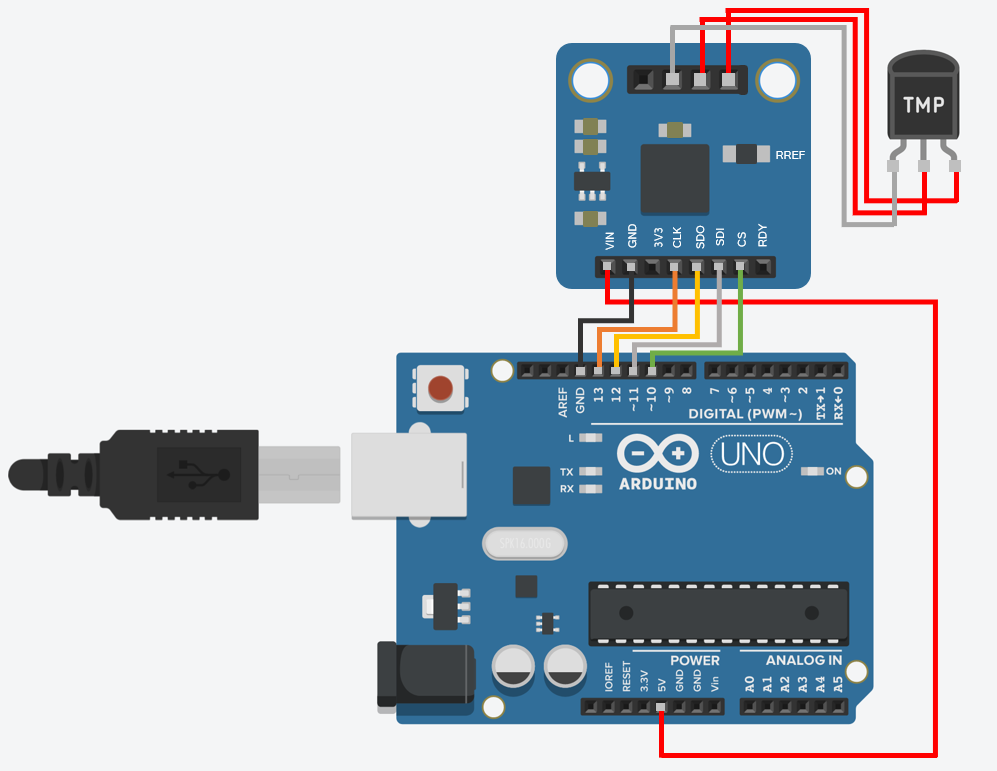}
    \caption{The Arduino setup} \label{fig:arduino}
 \end{subfigure}
    \caption{\textbf{The RTD temperature sensor and Arduino setup.} 
    (a) A 3-Wire RTD interface assumes shared parasitic lead resistance. The resistance of the platinum resistor $R_{Pt}=R_{AC} - R_{BC}$.
    (b) The RTD is connected directly to the MAX31865, and solder-pads are shorted to set the amplifier up in a 3-wire configuration. The Arduino supplies power to the amplifier. Additionally, digital pins are used to setup the serial peripheral interface communication. The selected digital pins on the Arduino used for clock, serial data in/out, and chip select connections are specified in software.
    }
\end{figure}

To measure the temperature of the lime as it comes to thermal equilibrium with the ambient air, we use a resistance temperature detector (RTD) sensor. Particularly, we use a PT1000, a high precision, 3-wire platinum RTD sensor with a nominal resistance $R_0=1000\,\Omega$ at 0\,$^\circ$C. As observed in Fig.~\ref{fig:rtd}, the PT1000 is constructed out of a platinum resistor connected in between 3 copper wires. As the temperature of the platinum resistor changes, so does its electrical resistance. Measuring the resistance between points A and C yields the resistance of the platinum resistor ($R_{Pt}$) plus the lead resistance of wire A and C. A 3-wire interface is used to mitigate the impact of parasitic resistance of the copper leads. The resistance between points B and C ($R_{BC}$) is subtracted from that of points A and C ($R_{AC}$) using an analog-to-digital converter to isolate the resistance of the platinum as $R_{Pt}$, which is then used to infer the corresponding temperature via the resistance-temperature relationship.

The relationship between the resistance of the platinum and its temperature $T > 0$ [$^\circ$C] is characterized by the Callendar-Van Dusen equation \cite{wu2018basic}:
\begin{equation}
    R_{Pt}(T) = 
        R_0 (1 + A T + B T^2), \label{eq:csv}
\end{equation}
with coefficients $A=3.9083\times10^{-3}$ [$\circ$C]$^{-1}$ and $B=–5.775\times 10^{-7}$ [$\circ$C]$^{-2}$ \cite{king2004rtd}. 
Software implementation details are available at \url{https://github.com/adafruit/Adafruit_MAX31865}.

In our experiment, the RTD sensor is paired with a thermocouple amplifier, an Adafruit MAX31865. This amplifier enables the observation of small changes in resistance at high fidelity using an on-board analog-to-digital converter with a 0.1\% reference resistor. The MAX31865 reads the resistance of the RTD sensor, then relays its findings to the micro-controller over a serial peripheral interface.

Onboard our micro-controller, an Arduino Uno, we signal the MAX31865 to capture a reading of the lime's temperature every 10 seconds. This data is communicated back to the Arduino. From here, the Arduino computes the temperature of the lime given the resistance of the RTD sensor through the solution to eqn.~\ref{eq:csv}. This computation is quick, and the temperature and wall-clock time are communicated to a connected computer over a universal asynchronous receiver/transmitter interface. The computer populates these readings into a comma-separated value file for further data processing. See Fig.~\ref{fig:arduino}.

\clearpage 

\subsection{Convergence diagnostics for the Markov chain Monte Carlo simulation}

\begin{figure}[h!]
    \centering
    \includegraphics[width=0.7\textwidth]{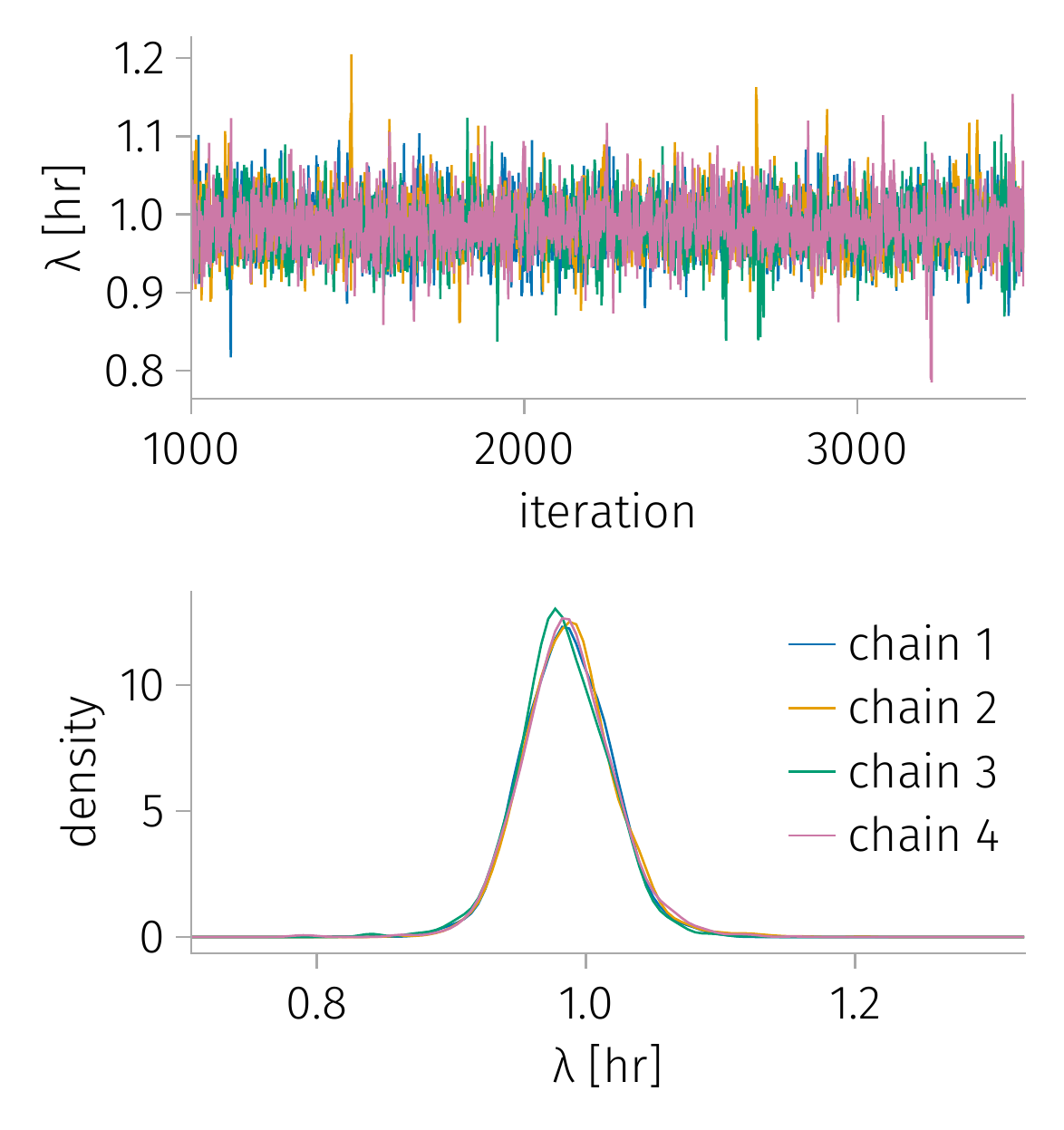}
    \caption{\textbf{Convergence diagnostics \cite{roy2020convergence} for our NUTS routine for inverse problem I.} The four colors correspond to four independent Markov chains. 
    (a) The trace plots show the parameter $\lambda_i$ at step $i$ of the Markov chain. Note, (i) approximately the same range of $\lambda$'s are explored in each chain and (ii) we see good mixing of the chain, in that local maxima and minima are frequently encountered presented. The iterations start at 1000 because the samples from the preceding iterations were discarded as ``burn-in''.
    (b) The empirical posterior distribution of $\Lambda$. Note, over the four independent chains, the posterior distribution is approximately consistent. 
    }
    \label{fig:convergence_diagnostics}
\end{figure}

\clearpage

\subsection{Residuals}

\begin{figure}[h!]
    \centering
    \includegraphics[width=0.7\textwidth]{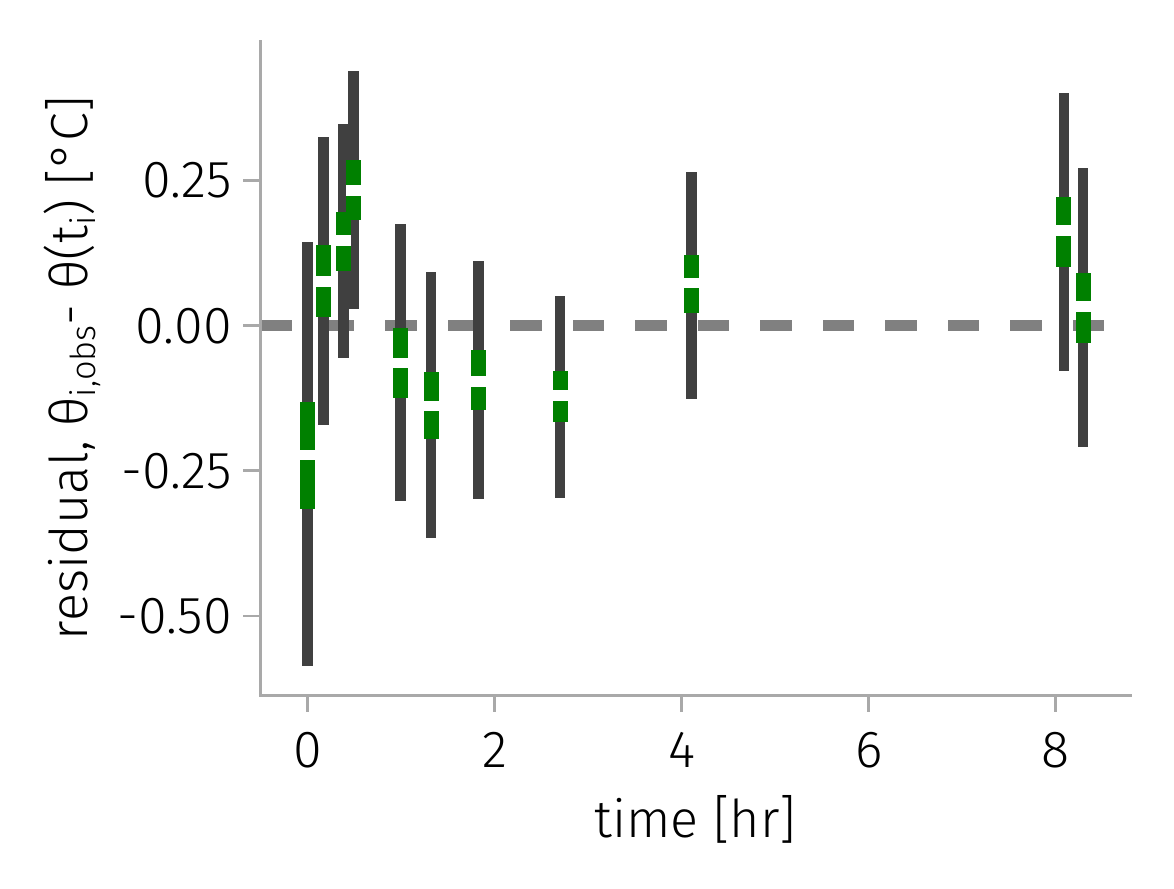}
    \caption{\textbf{Residuals.}
    For inverse problem I, we use a box plot to visualize the posterior distribution of each residual $\theta_{i, \text{obs}}-\theta(t_i; \Lambda, t_0=0, \Theta_0, \Theta^{\text{air}})$.
    }
    \label{fig:residuals}
\end{figure}

\clearpage

\subsection{A graphical solution to inverse problem IIa}
\begin{figure}[h!]
    \centering
    \includegraphics[width=0.7\textwidth]{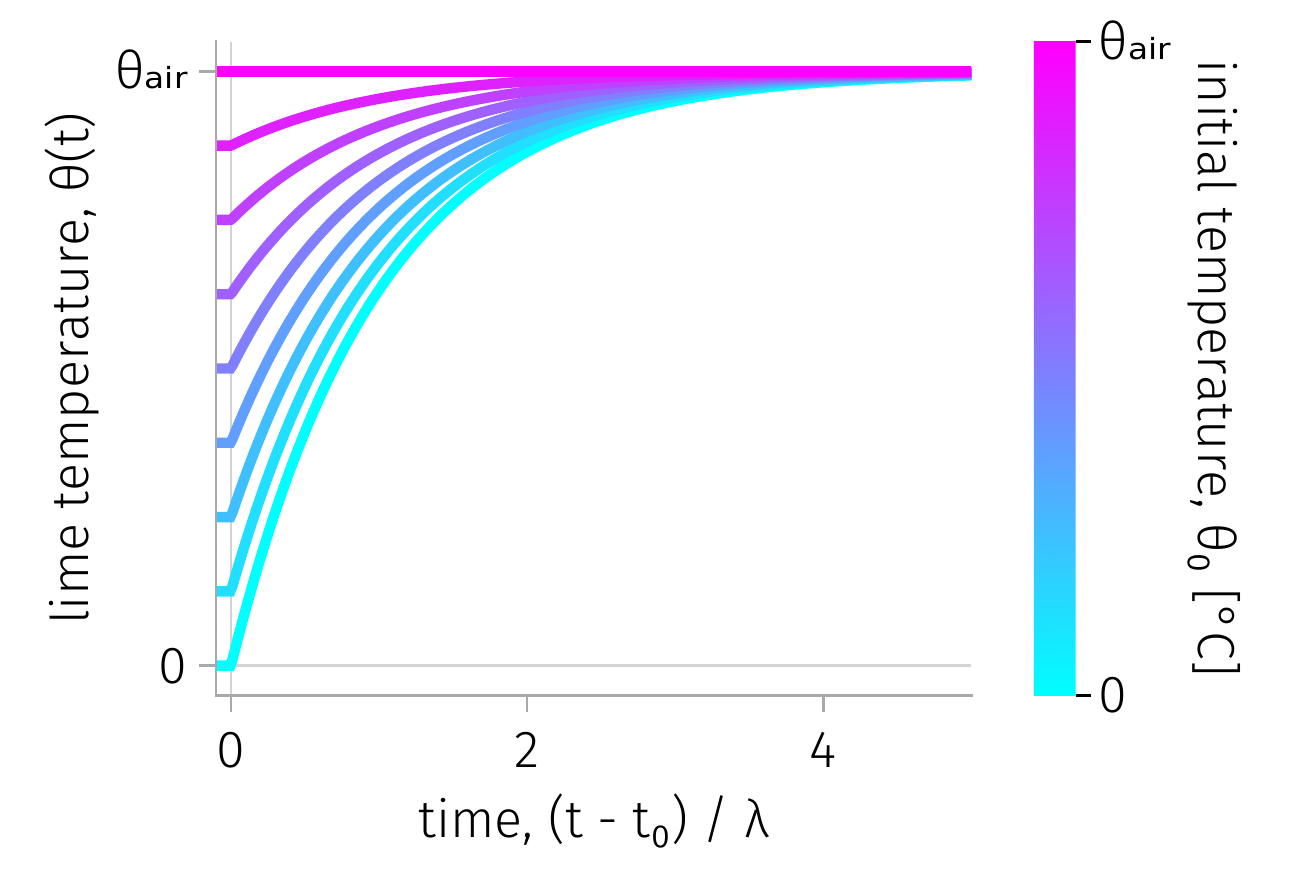}
    \caption{\textbf{A graphical solution to inverse problem IIa.}
    Trajectories of the lime temperature for different initial ($t=t_0$) temperatures $\theta_0$ at according to the model in eqn.~\ref{eq:model_soln}. 
    Given a measurement of the lime temperature, $\theta^\prime$, at a later time $t^\prime$, the predicted initial lime temperature $\theta_0$, ie.\ the classical solution to inverse problem IIa, follows from tracing the trajectory that passes through $(t^\prime, \theta^\prime)$ back in time to $t=t_0$.
    Visually, this inverse problem becomes ill-conditioned at large times, as a small perturbation in $\theta^\prime$ leads to a dramatically different backward trajectory. 
    }
    \label{fig:graphical_soln}
\end{figure}
\end{document}